\documentclass[journal]{IEEEtran}
\usepackage{amsmath,amsfonts}
\usepackage{algorithmic}
\usepackage{algorithm}
\usepackage{array}
\usepackage[caption=false,font=footnotesize,labelfont=rm,textfont=rm]{subfig}
\usepackage{textcomp}
\usepackage{stfloats}
\usepackage{url}
\usepackage{verbatim}
\usepackage{graphicx}
\usepackage{cite}
\usepackage{balance}
\usepackage{xcolor}
\usepackage{multirow}
\usepackage{multicol}
\usepackage{makecell}
\usepackage{tabularx}
\usepackage{booktabs}
\usepackage{titlesec}
\usepackage{amsmath}
\usepackage{amssymb}
\usepackage{amsthm}
\usepackage{bm}
\usepackage{bbm}
\usepackage{graphicx} % Required for inserting images
\usepackage{svg}

\newtheoremstyle{remarkbf}% 〈style name〉
  {3pt}{3pt}                 % space above / below
  {\normalfont}                 % body font (kept in italics)
  {}                         % indent amount
  {\bfseries}                % head font → bold
  {.}{ }{}                    % punctuation, spacing, theorem head spec

\theoremstyle{remarkbf}
\newtheorem{remark}{Remark}

\allowdisplaybreaks

\ifCLASSINFOpdf
\else
\fi
\begin{document}
%
% paper title
% Titles are generally capitalized except for words such as a, an, and, as,
% at, but, by, for, in, nor, of, on, or, the, to and up, which are usually
% not capitalized unless they are the first or last word of the title.
% Linebreaks \\ can be used within to get better formatting as desired.
% Do not put math or special symbols in the title.

\title{A Uniform Pilot and Data Payload Optimization Framework for OTFS-Based ISAC}

\author{Borui Du, Yumeng Zhang, Christos Masouros,~\IEEEmembership{Fellow,~IEEE}, Bruno Clerckx,~\IEEEmembership{Fellow,~IEEE}
        % <-this % stops a space

\thanks{Part of this work has been submitted to IEEE ICC 2026 for potential publication. Compared with the conference submission, this journal version provides complete derivations and a fully developed optimization framework, adds OTFS–OFDM capacity-lower-bound comparisons and throughput-based evaluations, and offers new analyses that yield clear OTFS-ISAC waveform design insights \cite{du_signal_2025}.}
    
\thanks{Borui Du and Christos Masouros are with the Department of Electronic and Electrical Engineering, University College London, London WC1E 6BT, U.K. (email:borui.du.24@ucl.ac.uk, c.masouros@ucl.ac.uk).}

\thanks{Yumeng Zhang is with the Department of Electronic and Computing Engineering, the Hong Kong University of Science and Technology, Hong Kong (email:eeyzhang@ust.hk).}

\thanks{Bruno Clerckx is with the Department of Electrical and Electronic Engineering, Imperial College London, London SW7 2AZ, U.K. (e-mail:
b.clerckx@imperial.ac.uk).}}
% note the % following the last \IEEEmembership and also \thanks - 
% these prevent an unwanted space from occurring between the last author name
% and the end of the author line. i.e., if you had this:
% 
% \author{....lastname \thanks{...} \thanks{...} }
%                     ^------------^------------^----Do not want these spaces!
%
% a space would be appended to the last name and could cause every name on that
% line to be shifted left slightly. This is one of those "LaTeX things". For
% instance, "\textbf{A} \textbf{B}" will typeset as "A B" not "AB". To get
% "AB" then you have to do: "\textbf{A}\textbf{B}"
% \thanks is no different in this regard, so shield the last } of each \thanks
% that ends a line with a % and do not let a space in before the next \thanks.
% Spaces after \IEEEmembership other than the last one are OK (and needed) as
% you are supposed to have spaces between the names. For what it is worth,
% this is a minor point as most people would not even notice if the said evil
% space somehow managed to creep in.

% The paper headers
\markboth{IEEE Transactions on Signal Processing}%
{Shell \MakeLowercase{\textit{et al.}}: Bare Demo of IEEEtran.cls for IEEE Journals}
% The only time the second header will appear is for the odd numbered pages
% after the title page when using the twoside option.
% 
% *** Note that you probably will NOT want to include the author's ***
% *** name in the headers of peer review papers.                   ***
% You can use \ifCLASSOPTIONpeerreview for conditional compilation here if
% you desire.

% If you want to put a publisher's ID mark on the page you can do it like
% this:
%\IEEEpubid{0000--0000/00\$00.00~\copyright~2015 IEEE}
% Remember, if you use this you must call \IEEEpubidadjcol in the second
% column for its text to clear the IEEEpubid mark.

% use for special paper notices
%\IEEEspecialpapernotice{(Invited Paper)}

% make the title area
\maketitle

% As a general rule, do not put math, special symbols or citations
% in the abstract or keywords.
\begin{abstract}
The orthogonal time frequency space (OTFS) signal is considered a promising solution for high-mobility wireless environments. It manages Doppler effects by utilizing delay-Doppler (DD) domain processing. However, the relatively long OTFS frame duration could introduce considerable sensing or communication latency when radar and communication are performed separately. By operating in a dual-functional radar and communication (DFRC) mode, the OTFS system performs sensing and data transmission simultaneously, thereby reducing the resulting latency. Nevertheless, the optimal OTFS DFRC signal strategy remains insufficiently explored. This paper investigates the optimal signal design for OTFS DFRC systems, focusing on pilot symbol design and data symbol power allocation. Specifically, we derive a channel capacity lower bound metric for communication that considers channel estimation errors in OTFS. For sensing, we derive an integrated sidelobe level (ISL), accounting for the randomness of the data symbols alongside the deterministic pilot symbols. Leveraging the above metrics, we formulate an optimization problem that balances radar and communication performance, and then solve it using an alternating optimization framework. We validate the proposed signal through numerical analysis and Monte Carlo simulations. Our analysis shows that OTFS DFRC enforces a deterministic pilot signal that is characterized by a concentrated peak in the DD domain, which furnishes a common structure in the DD domain facilitating sensing and channel estimation, with data multiplexed in other DD grids, thereby unifying sensing and communication within a single OTFS signal. Compared with conventional OTFS signals, the proposed OTFS DFRC signal expands the achievable sensing-communication performance region, delivering at least a 9.45 dB ISL suppression for sensing and a 4.82 dB signal-to-interference-plus-noise ratio gain for communication.
\end{abstract}

% Note that keywords are not normally used for peerreview papers.

% For peer review papers, you can put extra information on the cover
% page as needed:
% \ifCLASSOPTIONpeerreview
% \begin{center} \bfseries EDICS Category: 3-BBND \end{center}
% \fi
%
% For peerreview papers, this IEEEtran command inserts a page break and
% creates the second title. It will be ignored for other modes.
\IEEEpeerreviewmaketitle

\section{Introduction}
\subsection{Literature Review}
\looseness=-1
\IEEEPARstart{I}{ntegrated} sensing and communication (ISAC) is recognized as a key enabler for future wireless technologies, such as autonomous vehicle networks and smart cities~\cite{liu_joint_2020, cheng_integrated_2022}. The coexistence paradigm, an initial implementation of ISAC~\cite{wang_power_2019, liu_joint_2020}, suffers from significant interference suppression overhead. To enhance integration, research attention has increasingly shifted toward the dual-functional radar and communication (DFRC) framework, in which radar and communication functions are performed using a common signal, thereby improving hardware utilization, conserving time-frequency resources, and reducing scheduling overhead~\cite{liu_tutorial_2021}.

In DFRC research, waveform analysis and signal design are pivotal for exploiting the achievable performance limits of ISAC systems. Xu et al. investigate frequency modulated continuous wave (FMCW) for DFRC, providing an analysis of the resulting radar performance degradation and proposing a compensation strategy~\cite{xu_radar_2023}. In~\cite{zhang_input_2024}, Zhang et al. investigate orthogonal frequency-division multiplexing (OFDM) signal design in DFRC systems, and the results demonstrate that randomness benefits communication while signal determinism improves radar performance. In~\cite{li_mimo_ofdm_2025}, Li et al. propose a symbol-level precoding signal design for multi-input multi-output OFDM ISAC systems. However, the FMCW signal has limited communication performance, and OFDM signaling may not be effective in high‑mobility scenarios, as such mobility could induce severe inter‑carrier interference (ICI).

In addition to FMCW and OFDM signals, the orthogonal time frequency space (OTFS) waveform has demonstrated significant potential as a promising candidate for DFRC systems. Specifically, OTFS multiplexes data in the delay–Doppler (DD) domain, which yields slowly varying effective channel coefficients~\cite{rasheed_sparse_2020} and thus remains robust under large Doppler shifts. This DD-domain signal representation also enables more advanced and accurate channel estimation in doubly selective channels~\cite{wang_otfs_2021}, improving bit error rate (BER) during data detection~\cite{srivastava_bayesian_2021}. Since the introduction of OTFS in~\cite{hadani_orthogonal_2017}, numerous studies have shown that OTFS achieves excellent radar performance while simultaneously enhancing communication reliability in high-mobility scenarios~\cite{hadani_orthogonal_2017, gaudio_effectiveness_2020, compagnoni_otfs_2025}.
 
Considerable effort has been devoted to enhancing OTFS communication, particularly through pilot design. Wang et al.~\cite{wang_pilot_2021} propose a pilot design scheme for OTFS modulation by optimizing the measurement matrix structure formulated by the pilot symbols. Utilizing a heuristic optimization method, the proposed scheme significantly improves channel estimation while reducing pilot overhead. In contrast,~\cite{srivastava_bayesian_2021} and~\cite{zhao_sparse_2020} propose two other sophisticated pilot designs relying on Bayesian learning-based channel estimation. Their results show that the proposed designs outperform other compressed sensing channel estimation methods. Van der Werf et al. \cite{werf_optimal_2026} analyze the channel estimation performance and pilot arrangement, and derive the mean square error (MSE)-optimal pilot design. Besides pilot design, Xiang et al.~\cite{xiang_robust_2024} propose a non-orthogonal multiple access scheme-assisted OTFS-ISAC framework to optimally arrange power allocation for multi-user communication.

In addition to communication, early efforts have been made on OTFS signal design for radar applications\cite{correas_serrano_emerging_2025, karimian_sichani_otfs_2025}, with some extensions to DFRC systems~\cite{song_otfs_dfrc_2023}. Specifically, Karimian-Sichani et al. optimize the ambiguity function (AF) of OTFS signals by suppressing the sidelobe level in range estimation~\cite{karimian_sichani_otfs_2025}. Furthermore, Song et al.~\cite{song_otfs_dfrc_2023} extend the signal design to DFRC scenarios by introducing phase perturbation to suppress the peak sidelobe level of the transmitted OTFS signal, but at the cost of an increased communication symbol error rate.
Additionally, Zhang et al. propose a majorization-minimization-based framework that jointly optimizes the transmitted symbols and the receive matched filter in OTFS DFRC to minimize communication interference and suppress the integrated sidelobe level (ISL) of the transmitted signal~\cite{zhang_dual_functional_2024}. This approach enhances detection performance and improves the achievable rate of the designed signal.
Very recently, Song et al.~\cite{song_low_2025} investigate OTFS signal design and optimization for ISAC systems, which separates the pilot and data symbol designs for pure radar and communication functionalities and provides insightful analysis of the respective functionalities.

While these studies have substantially advanced our understanding of the OTFS waveform for DFRC applications, several critical issues remain insufficiently investigated. Firstly, from the communication side, the influence of imperfect channel state information (CSI) has rarely been systematically analyzed and considered for signal design in OTFS DFRC. Secondly, when assessing OTFS sensing performance—particularly in terms of ISL—existing studies have rarely distinguished the randomness of data symbols and the determinism of pilot symbols. Finally, most OTFS ISAC studies have focused predominantly on either the sensing or the communication aspect, leaving a comprehensive joint performance analysis absent. In this paper, we optimize the DFRC performance of the OTFS signal in high-mobility channels by manipulating both communication data power and pilot symbols. Our contributions are summarized as follows:
\begin{itemize}
    \item To the best of our knowledge, this is the first paper to establish a comprehensive optimization framework for OTFS DFRC systems, treating jointly channel estimation error, data transmission, and sensing performance, while also addressing the interaction between deterministic pilots and random data symbols.
    
    \item For optimization metrics, we derive a tractable channel capacity lower bound for OTFS data transmission that accounts for channel estimation errors. Taking into account both random data symbols and deterministic pilot symbols, we develop an ISL expression for radar performance using the whole OTFS signal frame, where data symbols are modeled as circularly symmetric complex Gaussian (CSCG) random variables.
    
    \item Building on the derived channel capacity lower bound and the ISL expression, we formulate an optimization problem with respect to (w.r.t.) communication data power and pilot symbols in the DD domain. Specifically, we first decompose the overall problem into two subproblems using alternating optimization (AO). By introducing auxiliary variables, the first AO subproblem is formulated as a convex problem w.r.t. the data power, and could be directly solved. For the second AO problem, we apply the alternating direction method of multipliers (ADMM) to determine the DD domain pilot symbols. Within the ADMM framework, the first ADMM subproblem is convexified using successive convex approximation (SCA), and the second ADMM subproblem could be directly formulated as a convex problem.
    
\end{itemize}
Furthermore, for clarity, we list some important insights as follows:
\begin{itemize}
    \item Building on our formulation and simulations, we demonstrate that OTFS enables a structurally unified DFRC design by enforcing a pilot signal that is characterized by a concentrated peak in the DD domain, which facilitates precise sensing and channel estimation, while allowing data to be multiplexed on other DD grids for information transmission. The rationale for the common peak-like pilot structure is that, for communication, it minimizes channel estimation MSE~\cite{werf_optimal_2026}; for sensing, it concentrates the AF mainlobe and suppresses sidelobes~\cite{chong_DD_2025}; thus, the optimized OTFS DFRC signal structurally unifies both sensing and communication within a single signal.
    \item This configuration not only achieves orthogonal placement of pilot and data symbols in the DD domain but also ensures that both symbol types, after the inverse symplectic finite Fourier transform (ISFFT), utilize all frequency and time resources in the time-frequency (TF) domain. Hence, the proposed OTFS signal effectively mitigates the typical DFRC trade-off seen in OFDM, whose TF-domain resources are either exclusively occupied by pilots or exclusively by data, necessitating careful design.
\end{itemize}

\subsection{Organization}
The rest of this paper is organized as follows. Section~\ref{sec:sys_mod} introduces the system model of the OTFS transmission system. The performance metrics for OTFS transmission are presented in Section~\ref{sec:Per_metr}, including a lower bound on channel capacity and the ISL of the transmitted signal’s AF. Section~\ref{sec:Opt_prob} then formulates and solves the optimization problem for the DFRC functionality of OTFS. Numerical results and simulations are presented in Section~\ref{sec:Num_res}. The conclusion of this paper is formulated in Section~\ref{sec:conclu}.

\subsection{Notations}
In this paper, matrices and vectors are denoted by upper-case bold and lower-case bold letters, e.g., \( \mathbf{A} \) and \( \mathbf{a} \), while \(\left[\mathbf{A}\right]_{i,j}\) represents the element in the \(i\)-th row and \(j\)-th column of the matrix \(\mathbf{A}\). \( \mathbb{C} \) and \( \mathbb{R} \) represent the sets of complex and real numbers, respectively. The operators \( (\cdot)^T \), \( (\cdot)^* \), and \( (\cdot)^H \) denote matrix transpose, conjugate, and conjugate transpose, respectively. The notation \( \mathcal{CN}(0, \mathbf{C}) \) represents a zero-mean CSCG distribution with covariance matrix \( \mathbf{C} \). The trace, vectorization, and Kronecker product operators are denoted as \( \operatorname{Tr}(\cdot) \), \( \operatorname{vec}(\cdot) \), and \( \otimes \), respectively. \( \operatorname{vec}^{-1}(\cdot) \) is the reshape function that transforms a vector into a matrix, and $\operatorname{diag}\left(\cdot\right)$ is the diagonal matrix operator. Expectation is represented by \( \varepsilon\{\cdot\} \). The real function is denoted as \(\operatorname{Re}\left(\cdot\right)\). The $N$-point normalized discrete Fourier matrix is denoted as $\mathbf{F}_N$. The identity matrix of dimension $N$ is denoted as \( \mathbf{I}_{N} \). $\mathbbm{1}_{(\cdot)}$ is the indicator function. $\|\cdot\|_2$ and $\|\cdot\|_F$ denote $l_2$ and Frobenius norms, respectively.

\section{System and Signal Model}
In this section, we describe the OTFS signal model, the channel propagation, and the received signals mathematically.
\label{sec:sys_mod}

\begin{figure*}[ht]
    \centering
    \includegraphics[width=0.8\textwidth]{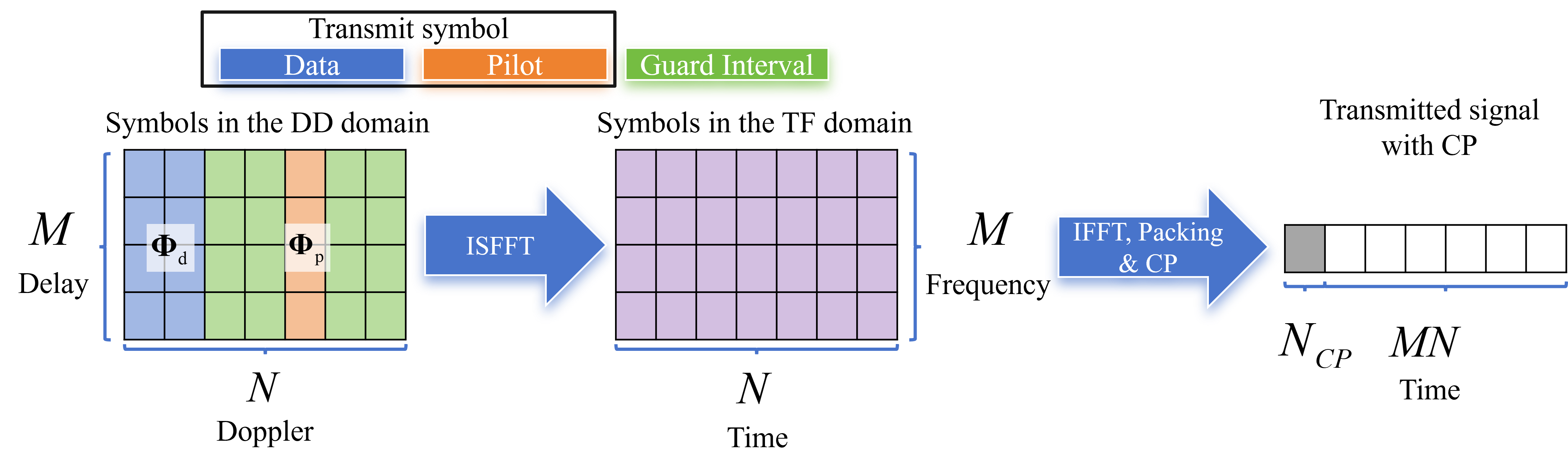}
    \caption{A toy example that illustrates the symbol arrangement (i.e., \(\mathbf{\Phi}_{\text{d}}\), \(\mathbf{\Phi}_{\text{p}}\)) and modulation process in the OTFS framework.}
    \label{fig:OTFS_mod}
\end{figure*}

\subsection{Transmitted Signal Model}
We first represent the OTFS signal in the DD domain. Each OTFS frame consists of \( M \) subcarriers and \( N \) time slots. Given a time slot duration of \( T \), the frequency subcarrier spacing is \( \Delta f = 1/T \), which corresponds to an OTFS bandwidth of \( B \triangleq M\Delta f \) and an OTFS duration of \( NT \).

We then denote the transmit symbol vector in the DD domain as \( \mathbf{x}_\text{DD} \in \mathbb{C}^{MN \times 1} \). In an OTFS system, as shown in Fig.~\ref{fig:OTFS_mod}, apart from the null-symbol guard space (green), the transmit symbols $\mathbf{x}_{\text{DD}}$ consist of two parts~\cite{werf_optimal_2026, xie_2025_bistatic}: (a) $K_p$ pilot symbols, denoted by $\mathbf{x}_{\text{DD,p}} \in \mathbb{C}^{K_p \times 1}$, which are used for channel estimation, and (b) $K_d$ data symbols, denoted by $\mathbf{x}_{\text{DD,d}} \in \mathbb{C}^{K_d \times 1}$, which carry random information. Additionally, we introduce arrangement matrices to map the pilot and data symbols into the corresponding DD resources in $\mathbf{x}_\text{DD}$, denoted by \( \mathbf{\Phi}_\text{p} \in \mathbb{R}^{MN \times K_p} \) and \( \mathbf{\Phi}_\text{d} \in \mathbb{R}^{MN \times K_d} \), respectively. Consequently, the OTFS communication symbol vector $\mathbf{x}_\text{DD}$ is expressed as follows
\begin{align}
    \mathbf{x}_\text{DD} = \mathbf{\Phi}_\text{d}\mathbf{x}_\text{DD,d} + \mathbf{\Phi}_\text{p}\mathbf{x}_\text{DD,p}, \label{equ:arrange_matrix}
\end{align}
where the data symbols are modeled as a CSCG codebook: $\mathbf{x}_\text{DD,d} \sim \mathcal{CN}\left(0,\, p_d\mathbf{I}_{K_d}\right)$, with variance $p_d$.

After arranging $\mathbf{x}_\text{DD,p}$ and $\mathbf{x}_\text{DD,d}$ in the DD domain, these symbols are transformed into the TF domain for OTFS modulation via the ISFFT, which is expressed as
\begin{subequations}
    \begin{align}
        {\mathbf{X}_\text{TF}} &= {{\mathbf{F}_M}{\mathbf{X}_\text{DD}}\mathbf{F}_N^H}, \\
        \text{ with } {\mathbf{X}_\text{DD}} &\triangleq \operatorname{vec}^{-1}\left({\mathbf{x}_\text{DD}} \right),
    \end{align}
\end{subequations}
where \( {\mathbf{X}_\text{DD}} \in {\mathbb{C}^{M \times N}} \) and \( {\mathbf{X}_\text{TF}} \in {\mathbb{C}^{M \times N}} \) represent the overall transmit symbols in the DD and TF domains, respectively.

We then transform $\mathbf{X}_\text{TF}$ into the delay-time (DT) domain for wireless propagation. After the inverse discrete Fourier transform, the signal ${\mathbf{S}_\text{DT}}$ in the DT domain is given by
\begin{align}
    {\mathbf{S}_\text{DT}} &= {\mathbf{P}_\text{tx}}\mathbf{F}_M^H{\mathbf{X}_\text{TF}} = {\mathbf{P}_\text{tx}}\mathbf{F}_M^H{\left({{\mathbf{F}_M}{\mathbf{X}_\text{DD}}\mathbf{F}_N^H}\right)},
\end{align}
where $\mathbf{P}_{\text{tx}}\in\mathbb{C}^{M\times M}$ denotes the diagonal pulse‑shaping matrix, which is set to the identity matrix for the rectangular‑pulse OTFS configuration, i.e., $\mathbf{P}_{\text{tx}}=\mathbf{I}_M$~\cite{raviteja_practical_2019}. To pack the entire frame for transmission, we vectorize $\mathbf{S}_{\text{DT}}$ as
\begin{equation}
    \mathbf{s} = \operatorname{vec}\left({{\mathbf{S}_\text{DT}}}\right) = \left({{{\mathbf{F}}_N^H} \otimes {{\mathbf{I}}_M}}\right) \mathbf{x}_\text{DD}.
\end{equation}
We then add a reduced cyclic prefix (CP) before transmission to mitigate inter-symbol interference. The transmit symbol vector in the DT domain, after adding CP, is expressed as follows
\begin{equation}
    {\mathbf{\widetilde{s}}} = \mathbf{\Gamma} {\mathbf{s}},
\end{equation}
where \( \mathbf{\Gamma} \in \mathbb{R}^{(MN + N_{\text{CP}}) \times MN} \) is the CP arrangement matrix, which is given by~\cite{zhang_dual_functional_2024}
\begin{subequations}
\begin{align}
    \mathbf{\Gamma} &=
    \left[\mathbf{\Gamma}_{\text{CP}}^{T}, \mathbf{I}_{MN}\right]^{T}, \\
    \text{with }[\mathbf{\Gamma}_{\text{CP}}]_{i,j} &= 
    \begin{cases}
        1, & \text{if } j = i + MN - N_{\text{CP}}\\
        0, & \text{otherwise}
    \end{cases},
\end{align}
\end{subequations}
where $N_\text{CP}$ denotes the length of the reduced CP, which is assumed to be larger than the maximum channel delay. As a consequence, \( {\mathbf{\widetilde{s}}} \in \mathbb{C}^{\left(MN + N_\text{CP}\right) \times 1} \).

\subsection{High-Mobility Channel Model}
To accommodate the Doppler effect, the wireless channel is modeled by a linear time-variant system. The channel effect on the OTFS signal could be represented using a matrix $\mathbf{H}\in\mathbb{C}^{MN \times MN}$~\cite{raviteja_practical_2019}, which is given by
\begin{subequations}
\begin{align}
  \mathbf{H}
    &= \sum_{i=1}^{P} \alpha_i\,\mathbf{\Pi}^{l_i}\,\mathbf{\Delta}^{k_i} \,, \\[6pt]
  \mathbf{\Pi}
    &= 
    \begin{bmatrix}
      0       & \cdots & 0      & 1 \\[2pt]
      1       & \cdots & 0      & 0 \\[2pt]
      \vdots  & \ddots & \vdots & \vdots \\[2pt]
      0       & \cdots & 1      & 0
    \end{bmatrix}, \\[6pt]
  \mathbf{\Delta}
    &= \operatorname{diag}\!\bigl(1,\;e^{j\frac{2\pi}{MN}},\;\dots,\;e^{j\frac{2\pi(MN-1)}{MN}}\bigr),
\end{align}
\end{subequations}
where $P$ denotes the number of propagation paths, and $\alpha_i$ is the complex gain of the $i$-th path. The matrix $\mathbf{\Pi}\in\mathbb{C}^{MN\times MN}$ is the cyclic-shift matrix, so that $\mathbf{\Pi}^{l_i}$ denotes a delay of $l_i$ for the $i$-th path. Likewise, $\mathbf{\Delta}\in\mathbb{C}^{MN\times MN}$ is the Doppler-shift matrix, and $\mathbf{\Delta}^{k_i}$ represents a Doppler shift of $k_i$ for the $i$-th path\footnote{This paper neglects fractional delays and Doppler shifts under the assumption that the bandwidth and frame duration are sufficiently large to provide enough resolution in the delay-Doppler domain~\cite{raviteja_practical_2019}.}. We assume the maximum delay and Doppler shift indices are denoted by \( L \) and \( Q \), respectively.

\subsection{Received Signal Model at the Communication Receiver}
After wireless channel propagation, the received signal after CP removal could be described as follows
\begin{equation}
    \mathbf{r} = \mathbf{H}\mathbf{s} + \mathbf{n},
\end{equation}
where $\mathbf{r}\in {\mathbb{C}}^{MN\times 1}$ is the received time-domain signal; $\mathbf{n}\in {\mathbb{C}}^{MN\times 1}$ is the additive white Gaussian noise (AWGN), with $\mathbf{n} \sim \mathcal{CN}\left(0, \sigma _{\bf{n}}^2\mathbf{I}_{MN}\right)$.
The received signal after CP removal is reshaped into the DT domain as follows
\begin{align}
    \mathbf{R} = {\operatorname{vec}^{ - 1}}\left( \mathbf{r} \right),
\end{align}
where \( \mathbf{R} \in \mathbb{C}^{M\times N} \) represents the received OTFS frame in the DT domain. The matched filtering and the discrete Fourier transform are applied to transform $\mathbf{R}$ from the DT to the TF domain, and the resultant signal in the TF domain is given by
\begin{align}
    \mathbf{Y}_\text{TF} &= {\mathbf{F}_M}{\mathbf{P}_\text{rx}}\mathbf{R},
\end{align}
where \( {\mathbf{P}_\text{rx}}\in \mathbb{C}^{M \times M} \) is the pulse-shaping matrix for the matched filtering. Similar to $\mathbf{P}_\text{tx}$, $\mathbf{P}_\text{rx}$ is assumed to be an identity matrix, i.e., $\mathbf{P}_{\text{rx}}=\mathbf{I}_M$\cite{raviteja_practical_2019}. Then, we apply the symplectic finite Fourier transform to transform $\mathbf{Y}_\text{TF}$ into the DD domain, which could be expressed as follows
\begin{align}
    \mathbf{Y}_\text{DD} &= {\mathbf{F}^H_M}\mathbf{Y}_\text{TF}{\mathbf{F}_N} = \mathbf{R}{\mathbf{F}_N},
\end{align}
whose vector form is given by
\begin{subequations}
    \begin{align}
        \mathbf{y}_\text{DD} &= \left( {{\mathbf{F}_N} \otimes {\mathbf{I}_M}} \right)\mathbf{H}\left({{{\mathbf{F}}_N^H} \otimes{{\mathbf{I}}_M}}\right) \mathbf{x}_\text{DD} + \mathbf{n}_\text{DD}\\
        &= \mathbf{H}_\text{DD} \mathbf{x}_\text{DD} + \mathbf{n}_\text{DD}, \label{equ:OTFS_sys}\\
        \text{ with }\mathbf{H}_\text{DD} &\triangleq \sum\limits_{i = 1}^P \alpha_i \left( \mathbf{F}_N \otimes \mathbf{I}_M \right){\mathbf{\Pi}}^{l_i} {\mathbf{\Delta}}^{k_i} \left(\mathbf{F}_N^H \otimes \mathbf{I}_M \right), \label{equ:h2H}
    \end{align}
\end{subequations} where \( \mathbf{n}_\text{DD} \triangleq \left( {{\mathbf{F}_N} \otimes {\mathbf{I}_M}} \right)\mathbf{n} \), and $\mathbf{H}_\text{DD}$ is the effective channel in the DD domain that characterizes the delay-Doppler channel effects and OTFS modulation.

As counterparts to $\mathbf{\Phi}_{\text{p}}$ and $\mathbf{\Phi}_{\text{d}}$ in (\ref{equ:arrange_matrix}), selection matrices \( \mathbf{\Psi}_\text{p} \in \mathbb{R}^{MN \times R_p} \) and \( \mathbf{\Psi}_\text{d} \in \mathbb{R}^{MN \times R_d} \) are applied to extract the pilot and data symbols from the DD domain, respectively. To decouple the pilot and data symbols, we assume that\cite{werf_optimal_2026}
\begin{equation}
\mathbf{\Psi}^H_\text{p} \mathbf{H}_\text{DD} \mathbf{\Phi}_\text{d} = 0, \quad
\mathbf{\Psi}^H_\text{d} \mathbf{H}_\text{DD} \mathbf{\Phi}_\text{p} = 0, \label{equ:decoup_assump}
\end{equation}
which ensures a guard interval (GI) between pilot and data symbols and prevents mutual interference.

Therefore, the received pilot symbols are given by
\begin{subequations}
\begin{align}
     \mathbf{y}_\text{DD,p} &= \mathbf{\Psi}^H_\text{p}\mathbf{y}_\text{DD} = \mathbf{\Psi}^H_\text{p} \mathbf{H}_\text{DD} \mathbf{\Phi}_\text{p}\mathbf{x}_\text{DD,p} + \mathbf{\Psi}^H_\text{p}\mathbf{n}_\text{DD}\\
     &= \mathbf{H}_\text{p}\mathbf{x}_\text{DD,p} + \mathbf{n}_\text{p}, \label{equ:pilot_DD}
\end{align}
\end{subequations} where \(\mathbf{\Psi}^H_\text{p} \mathbf{H}_\text{DD}\mathbf{\Phi}_\text{d}\mathbf{x}_\text{DD,d}=0\), \(\mathbf{H}_\text{p}\triangleq\mathbf{\Psi}^H_\text{p} \mathbf{H}_\text{DD} \mathbf{\Phi}_\text{p}\), and \(\mathbf{n}_\text{p}\triangleq\mathbf{\Psi}^H_\text{p}\mathbf{n}_\text{DD}\). \( \mathbf{y}_\text{DD,p} \in \mathbb{C}^{R_p\times 1} \) denotes the received pilot symbols, and $\mathbf{n}_\text{p}$ represents the corresponding noise. Similarly, the received data symbols are given by
\begin{subequations}
\begin{align}
     \mathbf{y}_\text{DD,d} &= \mathbf{\Psi}^H_\text{d} \mathbf{H}_\text{DD} \mathbf{\Phi}_\text{d}\mathbf{x}_\text{DD,d} + \mathbf{\Psi}^H_\text{d}\mathbf{n}_\text{DD}\\
     &= \mathbf{H}_\text{d}\mathbf{x}_\text{DD,d} + \mathbf{n}_\text{d}, \label{equ:OTFS_sys_data}
\end{align}
\end{subequations}
where \(\mathbf{H}_\text{d}\triangleq \mathbf{\Psi}^H_\text{d} \mathbf{H}_\text{DD} \mathbf{\Phi}_\text{d}\), and \(\mathbf{n}_\text{d}\triangleq\mathbf{\Psi}^H_\text{d}\mathbf{n}_\text{DD}\). \( \mathbf{y}_\text{DD,d} \in \mathbb{C}^{R_d\times 1} \) denotes the received data symbols, and $\mathbf{n}_\text{d}$ represents the noise in the data symbol region.

Given pilot symbols $\mathbf{x}_\text{DD,p}$, we need to recover the channel matrix $\mathbf{H}_\text{p}$. Hence, we rewrite (\ref{equ:pilot_DD}) as follows
\begin{subequations}
\begin{align}
     \mathbf{y}_\text{DD,p} &= \bm{\Omega}_\text{DD,p}\mathbf{h} + \mathbf{n}_\text{p}, \label{equ:pilot_Omega_h}\\ 
     \text{ with }
    {\bm{\Omega} _{\text{DD,p}}} &= \left[ {\bm{\omega} _{0,\text{DD,p}}^0, \bm{\omega} _{1,\text{DD,p}}^0, \cdots, \bm{\omega} _{L,\text{DD,p}}^{Q}} \right] \in \mathbb{C}^{R_p \times K_{h}},
\end{align}
\end{subequations} where \( \mathbf{h}\in\mathbb{C}^{K_{h}\times 1} \), with \( K_{h} \triangleq (L + 1)(Q + 1) \), is the channel response vector that follows a CSCG distribution of $\mathcal{CN}\left(0, \sigma^2_{\mathbf{h}}\right)$. The position of each entry represents the corresponding delay and Doppler shift. In addition, $\bm{\Omega} _{\text{DD,p}}$ is the pilot dictionary matrix, whose column vectors \( \bm{\omega} _{i,\text{DD,p}}^j \in \mathbb{C}^{R_p\times 1} \) are given by
\begin{equation}
    \bm{\omega} _{i,\text{DD,p}}^j = \mathbf{\Psi}^H_\text{p} \left( {\mathbf{F}_N \otimes {\mathbf{I}_M}} \right) {\mathbf{\Pi}}^i {\mathbf{\Delta}}^j \left( {\mathbf{F}^H_N \otimes {\mathbf{I}_M}} \right) \mathbf{\Phi}_\text{p} \mathbf{x}_{\text{DD,p}},
\end{equation}
which contain pilot arrangement information from $\mathbf{\Phi}_\text{p}$ and $\mathbf{\Psi}_\text{p}$, pilot symbol information from $\mathbf{x}_\text{DD,p}$, path delay information of $i$, and path Doppler information of $j$.

\section{Performance Metrics}
\label{sec:Per_metr}
In this section, we express the channel capacity lower bound for the communication metric that accounts for channel estimation error and derive a looser lower bound for tractability. Additionally, we present the radar ISL metric, which incorporates the Gaussian-distributed data symbols along with deterministic pilot symbols. These metrics are essential for the optimization methods discussed in Section~\ref{sec:Opt_prob}.

\subsection{Channel Capacity Lower Bound}
Given (\ref{equ:pilot_Omega_h}), the linear minimum mean square error (LMMSE) channel estimate could be expressed as~\cite{srivastava_bayesian_2021}
\begin{equation}
    \mathbf{\widehat h}_\text{LMMSE} = \left(\bm{\Omega}^H_\text{DD,p}\mathbf{C}^{-1}_{\mathbf{n}_\text{p}}\bm{\Omega}_\text{DD,p} + \mathbf{C}^{-1}_\mathbf{h}\right)^{-1}\bm{\Omega}^H_\text{DD,p}\mathbf{C}^{-1}_{\mathbf{n}_\text{p}}\mathbf{y}_\text{DD,p},
\end{equation}
where \( \mathbf{C}_{\mathbf{n}_\text{p}} \) is the noise covariance matrix of the pilot region. The estimated channel matrix \( \widehat{\mathbf{H}}_\text{DD} \) is then obtained from (\ref{equ:h2H}) using $\mathbf{\widehat h}_\text{LMMSE}\in\mathbb{C}^{K_h \times 1}$. Also, we denote \( \widehat{\mathbf{H}}_\text{d} \triangleq \mathbf{\Psi}^H_\text{d} \widehat{\mathbf{H}}_\text{DD}\mathbf{\Phi}_\text{d}\) and \( \widehat{\mathbf{H}}_\text{p} \triangleq \mathbf{\Psi}^H_\text{p} \widehat{\mathbf{H}}_\text{DD}\mathbf{\Phi}_\text{p}\) as the estimated channel matrices for data and pilot symbols, respectively. Therefore, $\mathbf{y}_\text{DD,d}$ could be rewritten as follows
\begin{align}
     \mathbf{y}_\text{DD,d} &= \widehat{\mathbf{H}}_\text{d}\mathbf{x}_\text{DD,d} + \left(\mathbf{H}_\text{d} -\widehat{\mathbf{H}}_\text{d}\right)\mathbf{x}_\text{DD,d} + \mathbf{n}_\text{d} \\ 
     &= \widehat{\mathbf{H}}_\text{d}\mathbf{x}_\text{DD,d} + \mathbf{v}_\text{DD,d}, \label{equ:OTFS_sys_data_est_error}
\end{align}
where $\mathbf{v}_\text{DD,d}\buildrel \Delta \over=\left(\mathbf{H}_\text{d} -\widehat{\mathbf{H}}_\text{d}\right)\mathbf{x}_\text{DD,d} + \mathbf{n}_\text{d}$ is the effective noise vector accounting for channel estimation error and AWGN.
Based on the received signal model in (\ref{equ:OTFS_sys_data_est_error}), the channel capacity lower bound for OTFS is given by~\cite{xiaoli_ma_optimal_2003}
\begin{equation}
    C \ge \frac{f_{\text{CP}}}{MN} \varepsilon_{\mathbf{H}_\text{d}} \left[\log \det \left(\mathbf{I}_{R_d} + {p_d} \mathbf{C}^{-1}_{\mathbf{v}_\text{DD,d}} \mathbf{\widehat H}_\text{d} \mathbf{\widehat H}^H_\text{d}\right)\right], \label{equ:OTFS_channel_capacity_lower_bound}
\end{equation}
where \( f_{\text{CP}} \buildrel \Delta \over= \frac{MN}{MN+N_\text{CP}} \) accounts for the CP overhead. As discussed in \cite{xiaoli_ma_optimal_2003}, this capacity lower bound is derived under Gaussian signaling and a Gaussian channel model, which maximize the entropy of the transmitted symbols (yielding the capacity) and the error term (yielding the lower bound). Hence, it constitutes a valid lower bound on the channel capacity.

$\mathbf{C}_{\mathbf{v}_\text{DD,d}}$ denotes the covariance of the effective error vector $\mathbf{v}_\text{DD,d}$, which could be written as follows
\begin{subequations}
\begin{align}
    \mathbf{C}_{\mathbf{v}_\text{DD,d}}
    &= \varepsilon \left[\mathbf{v}_\text{DD,d}\mathbf{v}_\text{DD,d}^H\right] \\
    &= p_d\varepsilon \left[ (\mathbf{H}_\text{d}-\mathbf{\widehat H}_\text{d})(\mathbf{H}_\text{d}-\mathbf{\widehat H}_\text{d})^H \right] + \sigma^2_{\mathbf{n}}\mathbf{I}_{R_d} \label{equ:CvDD_H}\\
    &= p_d\widetilde{{\bm{\Omega}}}_\text{DD,d} \left\{ {\varepsilon\left[(\mathbf{h} - \widehat{\mathbf{h}})(\mathbf{h} - \widehat{\mathbf{h}})^H \right] \otimes {\mathbf{I}_{K_d}}}\right\} \widetilde{{\bm{\Omega}}}_\text{DD,d}^H \notag\\
    &\quad+ \sigma^2_{\mathbf{n}}\mathbf{I}_{R_d} \\
    &\triangleq p_d\widetilde{{\bm{\Omega}}}_\text{DD,d} \left( {\mathbf{C}_{{{{\mathbf{\epsilon}}}}} \otimes {\mathbf{I}_{K_d}}}\right) \widetilde{{\bm{\Omega}}}_\text{DD,d}^H + \sigma^2_{\mathbf{n}}\mathbf{I}_{R_d},\label{equ:CvDD}
\end{align}
\begin{align}
    \text{with }\widetilde{{\bm{\Omega}}}_\text{DD,d} &= \left[ \widetilde{{\bm{\Omega}}}_{0,\text{DD,d}}^0, \widetilde{{\bm{\Omega}}}_{1,\text{DD,d}}^{0}, \cdots, \widetilde{{\bm{\Omega}}}_{{L},\text{DD,d}}^{Q} \right], \label{equ:Omega_tilde_c}\\
    \text{and }{\widetilde {\bm{\Omega}}}_{i,\text{DD,d}}^j &= \mathbf{\Psi}^H_\text{d}\left( {\mathbf{F}_N \otimes {\mathbf{I}_M}} \right){{{\bm{\Pi}}} ^i}{\bm{\Delta}}^j\left( {{\mathbf{F}_N^H} \otimes {\mathbf{I}_M}} \right)\mathbf{\Phi}_\text{d},
\end{align}
\end{subequations}
where each submatrix $\widetilde{\bm{\Omega}}^j_{i,\text{DD,d}}\in \mathbb{C}^{R_d\times K_d}$ is determined once the channel parameters and symbol arrangement ($K_h, R_d, K_d$) are specified. $\mathbf{C}_{{{{\mathbf{\epsilon}}}}} \in \mathbb{C}^{K_h \times K_h}$ is the estimation error covariance matrix of $\widehat{\mathbf{h}}$, which is defined as $\mathbf{C}_{\epsilon}\triangleq\varepsilon\left[(\mathbf{h} - \widehat{\mathbf{h}})(\mathbf{h} - \widehat{\mathbf{h}})^H \right]$. 
Using the LMMSE estimator, $\mathbf{C}_{{{{\mathbf{\epsilon}}}}}$ could be expressed as 
\begin{align}
{\mathbf{C}_{{\epsilon}}} = {\mathbf{C}_{{\mathbf{h}}}} - {\mathbf{C}_{{\mathbf{\widehat h}}}} = \sigma_{\mathbf{h}}^2 \left(\mathbf{I}_{K_{h}} + \frac{\sigma_{\mathbf{h}}^2}{\sigma_{\mathbf{n}}^2}{{\bm{\Omega}}^H _\text{DD,p}}{\bm{\Omega}} _\text{DD,p}\right)^{-1}, \label{equ:C_eps}
\end{align}
whose details are given in the Appendix. By substituting (\ref{equ:C_eps}) into (\ref{equ:CvDD}), the channel capacity lower bound of (\ref{equ:OTFS_channel_capacity_lower_bound}) is reformulated as (\ref{equ:OTFS_channel_capacity_lower_bound_mat}), with the inner term given in (\ref{equ:M_otfs}), both found at the top of this page.

\begin{figure*}[ht]
    \begin{equation}
        C \geq \frac{f_{\text{CP}}}{MN} \varepsilon \left[\log \det \left(\mathbf{I}_{R_d} + \underbrace{p_d\left(p_d \widetilde{{\bm{\Omega}}}_\text{DD,d} \left( {\mathbf{C}_{{{{\mathbf{\epsilon}}}}} \otimes {\mathbf{I}_{K_d}}}\right) \widetilde{{\bm{\Omega}}}_\text{DD,d}^H + \sigma^2_{\mathbf{n}}\mathbf{I}_{R_d}\right)^{-1}}_{\triangleq \mathbf{M}_\text{OTFS}} \mathbf{\widehat H}_\text{d} \mathbf{\widehat H}_\text{d}^H\right)\right] \label{equ:OTFS_channel_capacity_lower_bound_mat}
    \end{equation}
    \hrule
\end{figure*}
\begin{figure*}
    \begin{equation}
        \mathbf{M}_\text{OTFS} = p_d\left(p_d \widetilde {\bm{\Omega}}_{\text{DD,d}} \left( {\sigma_{\mathbf{h}}^2 \left(\mathbf{I}_{K_{h}} + \frac{\sigma_{\mathbf{h}}^2}{\sigma_{\mathbf{n}}^2}{{\bm{\Omega}}^H _\text{DD,p}}{\bm{\Omega}} _\text{DD,p}\right)^{-1} \otimes {\mathbf{I}_{K_d}}}\right) \widetilde {\bm{\Omega}}^H_{\text{DD,d}} + \sigma^2_\mathbf{n}\mathbf{I}_{R_d}\right)^{-1} \label{equ:M_otfs}
    \end{equation}
    \hrule
\end{figure*}

\begin{figure*}
    \begin{subequations}
        \begin{align}
            \mathbf{C}_{\mathbf{v}_\text{DD,d}} &\mathop \preceq {p_d} \varepsilon \left\{ \operatorname{Tr}\left[(\mathbf{h} - \widehat{\mathbf{h}})(\mathbf{h} - \widehat{\mathbf{h}})^H \right]\right\}\mathbf{I}_{R_d} + \sigma^2_\mathbf{n} \mathbf{I}_{R_d} \label{equ:tr_app}\\
            &= {p_d} \operatorname{Tr} \left\{ \mathbf{C}_{\epsilon} \right\}\mathbf{I}_{R_d} + \sigma^2_\mathbf{n} \mathbf{I}_{R_d}\\
            &= 
            \Bigg[ 
            p_d \operatorname{Tr} \Bigg( 
            \sigma_{\mathbf{h}}^2 
            \left( 
            \mathbf{I}_{K_h} + 
            \frac{\sigma_{\mathbf{h}}^2}{\sigma_{\mathbf{n}}^2} 
            \bm{\Omega}^H_\text{DD,p} \bm{\Omega}_\text{DD,p} 
            \right)^{-1} 
            \Bigg) 
            + \sigma^2_\mathbf{n} 
            \Bigg] \mathbf{I}_{R_d}
        \label{equ:tr_app_result_specific}
        \end{align}
    \end{subequations}
    \hrule
\end{figure*}
Given the covariance matrix $\mathbf{C}_{\epsilon}$ in (\ref{equ:C_eps}), we could upper-bound (\ref{equ:CvDD_H}) by a tractable expression in (\ref{equ:tr_app_result_specific})~\cite{werf_optimal_2026}. By combining (\ref{equ:OTFS_channel_capacity_lower_bound}) with (\ref{equ:tr_app_result_specific}), a looser but tractable capacity lower bound is given by (\ref{equ:OTFS_channel_capacity_lower_bound_scalar}) at the top of the next page, which, although sacrificing a certain degree of accuracy compared with the original capacity lower bound in (\ref{equ:OTFS_channel_capacity_lower_bound_mat}), allows tractable optimization in the following sections.
\begin{figure*}[ht]
    \begin{equation}
        C \ge \frac{f_\text{CP}}{MN} \varepsilon \left\{\log \det \left[\mathbf{I}_{R_d} + \underbrace{\frac{p_d}{\sigma^2_{\mathbf{n}}} \left[ \frac{p_d}{\sigma^2_{\mathbf{n}}} \operatorname{Tr} \left( \sigma_{\mathbf{h}}^2 \left(\mathbf{I}_{K_{h}} + \frac{\sigma_{\mathbf{h}}^2}{\sigma_{\mathbf{n}}^2}{{\bm{\Omega}}^H _\text{DD,p}}{\bm{\Omega}} _\text{DD,p}\right)^{-1}\right) + 1\right]^{-1}}_\text{Scalar} \mathbf{\widehat H}_\text{d} \mathbf{\widehat H}^H_\text{d}\right]\right\} \label{equ:OTFS_channel_capacity_lower_bound_scalar}
    \end{equation}
    \hrule
\end{figure*}
\begin{figure*}[ht]
\begin{align}
    \text{SINR} &\triangleq 
    \frac{p_d}{\sigma^2_{\mathbf{n}}} \left[ \frac{p_d}{\sigma^2_{\mathbf{n}}} \operatorname{Tr} \left( \sigma_{\mathbf{h}}^2 \left(\mathbf{I}_{K_{h}} + \frac{\sigma_{\mathbf{h}}^2}{\sigma_{\mathbf{n}}^2}{{\bm{\Omega}}^H _\text{DD,p}}{\bm{\Omega}} _\text{DD,p}\right)^{-1}\right) + 1\right]^{-1}
\label{equ:opt_SINR}
\end{align}
\hrule
\end{figure*}

From \eqref{equ:OTFS_channel_capacity_lower_bound_scalar}, the scalar (as marked) inside the logarithm dominates the capacity lower bound once channel estimation error is considered. We define this scalar term as the signal-to-interference-plus-noise ratio (SINR) in \eqref{equ:opt_SINR} at the top of the next page~\cite{werf_optimal_2026}, and use it as the communication metric. Moreover, \(\text{SINR}\) is concave in \(p_d\) and therefore suitable to employ in optimization.

\subsection{Integrated Sidelobe Level of Ambiguity Function}
In terms of the radar metric, we characterize the sensing performance based on the AF properties of the transmitted signal, including both pilots and data symbols, namely, the ISL. The cross-correlation of the signal induced by a target at delay bin $l$ and Doppler bin $k$ can be written as follows
\begin{align}
    f_{lk} &= \mathbf{x}_\text{DD}^H \,\mathbf{A}_{lk}\,\mathbf{x}_\text{DD},
\end{align}
where 
\begin{align}
    \mathbf{A}_{lk}
    &= \bigl(\mathbf{F}_{N}\otimes \mathbf{I}_{M}\bigr)\,
       \mathbf{\Gamma}^H\,\mathbf{J}_{l}\,\mathbf{D}_{k}\,\mathbf{\Gamma}\,
       \bigl(\mathbf{F}_{N}^H\otimes \mathbf{I}_{M}\bigr)
\end{align} is a constant matrix. Moreover, 
\begin{align}
    \mathbf{J}_{l} &\in \mathbb{C}^{(MN + N_\text{CP})\times (MN + N_\text{CP})} \nonumber \\
    &= \begin{bmatrix}
        \mathbf{0}_{\,l\times (MN + N_\text{CP}-l)} & \mathbf{0}_{\,l\times l} \\
        \mathbf{I}_{\,MN + N_\text{CP}-l} & \mathbf{0}_{\,(MN + N_\text{CP}-l)\times l}
       \end{bmatrix}
\end{align} is the linear delay matrix with $\mathbf{J}_{-l}=\mathbf{J}^T_{l}$, and 
\begin{equation}
    \mathbf{D}_{k}
    = \operatorname{diag}\bigl(1,\,
      e^{-\frac{j2\pi k}{MN + N_\text{CP}}},\,
      \ldots,\,
      e^{-\frac{j2\pi (MN + N_\text{CP}-1)k}{MN + N_\text{CP}}}\bigr)
\end{equation} is the Doppler‐shift matrix for the received signal. Thus, the ISL is expressed as
\begin{subequations}
    \label{equ:ISL}
    \begin{align}
    \text{ISL} &= \varepsilon\left[\sum\limits_{ {l,k} } {{{\left| {{f_{lk}}} \right|}^2}}\right] = \varepsilon\left[\sum\limits_{{l,k}} \left(\mathbf{x}^H_\text{DD}\mathbf{A}_{lk}\mathbf{x}_\text{DD} \mathbf{x}^H_\text{DD}\mathbf{A}^H_{lk}\mathbf{x}_\text{DD}\right)\right] \\
    &= \sum\limits_{l,k} \Bigg[\mathbf{x}_\text{DD,p}^H \mathbf{A}_{\text{p},lk} \mathbf{x}_\text{DD,p} \mathbf{x}_\text{DD,p}^H \mathbf{A}_{\text{p},lk}^H \mathbf{x}_\text{DD,p}
     \notag\\
    &\quad\quad\quad\;+p_d^2 \big( a_{lk} + \left| b_{lk} \right|^2 \big) + p_d \mathbf{x}_\text{DD,p}^H \mathbf{B}_{lk} \mathbf{x}_\text{DD,p} \notag\\
    &\quad\quad\quad\;+ 2p_d \operatorname{Re} \big( b_{lk} \mathbf{x}_\text{DD,p}^H \mathbf{A}_{\text{p},lk}^H \mathbf{x}_\text{DD,p} \big) \Bigg], \label{equ:ISL_simp}\\
    \text{with }&\mathbf{B}_{lk} \triangleq \mathbf{A}_{\text{p,d},lk} \mathbf{A}_{\text{d,p},lk} + \mathbf{A}_{\text{p,d},lk} \mathbf{A}_{\text{p,d},lk}^H  + \mathbf{A}_{\text{d,p},lk}^H \mathbf{A}_{\text{d,p},lk} \notag \\ 
    &\quad\quad\;\;+ \mathbf{A}_{\text{d,p},lk}^H \mathbf{A}_{\text{p,d},lk}^H,
    \end{align}
\end{subequations} where $l=-\widehat{L}, \cdots, \widehat{L}$ and $k=-\widehat{Q}, \cdots, \widehat{Q}$, excluding $(l,k)=(0,0)$, denote the candidate delay and Doppler bins considered in the ISL calculation.
For brevity, the notation of~(\ref{equ:ISL}) is summarized in Table~\ref{tab:constants}.
\begin{table}[ht]
    \centering
    \caption{Definitions of scalar and matrix constants.}
    \renewcommand{\arraystretch}{1.5}
    \begin{tabular}{c|l}
        \hline
        Symbol & Definition \\
        \hline
        $a_{lk}$ & $\operatorname{Tr} \big( \mathbf{\Phi}_\text{d}^H \mathbf{A}_{lk} \mathbf{\Phi}_\text{d} \mathbf{\Phi}_\text{d}^H \mathbf{A}_{lk}^H \mathbf{\Phi}_\text{d} \big)$ (scalar constant) \\
        $b_{lk}$ & $\operatorname{Tr} \big( \mathbf{\Phi}_\text{d}^H \mathbf{A}_{lk} \mathbf{\Phi}_\text{d} \big)$ (scalar constant) \\
        $\mathbf{A}_{\text{p},lk}$ & $\mathbf{\Phi}_\text{p}^H \mathbf{A}_{lk} \mathbf{\Phi}_\text{p}$ (matrix constant) \\
        $\mathbf{A}_{\text{p,d},lk}$ & $\mathbf{\Phi}_\text{p}^H \mathbf{A}_{lk}^H \mathbf{\Phi}_\text{d}$ (matrix constant) \\
        $\mathbf{A}_{\text{d,p},lk}$ & $\mathbf{\Phi}_\text{d}^H \mathbf{A}_{lk}^H \mathbf{\Phi}_\text{p}$ (matrix constant) \\
        \hline
    \end{tabular}
    \label{tab:constants}
\end{table}

\begin{remark}
In the denominator of (\ref{equ:opt_SINR}), the channel estimation error is determined by ${{\bm{\Omega}}^H_{\text{DD,p}}}{\bm{\Omega}}_{\text{DD,p}}$ and $p_d$. The matrix ${{\bm{\Omega}}^H_{\text{DD,p}}}{\bm{\Omega}}_{\text{DD,p}}$ is governed by the pilot symbols and also composes the AF (the first term in (\ref{equ:ISL_simp})). Thus, the two metrics are consistent in terms of the pilot structure characterized by ${{\bm{\Omega}}^H_{\text{DD,p}}}{\bm{\Omega}}_{\text{DD,p}}$. In particular, its diagonal elements correspond to the AF mainlobe, whereas the off-diagonal elements represent the AF sidelobes\footnote{A rigorous equivalence can be established by reformulating the AF under a circular-delay model (sensing with CP) \cite{liu_cp_2025}. Under the present linear-delay model (sensing without CP), this correspondence is not exact but exhibits similar behavior.}. Moreover, when ${{\bm{\Omega}}^H_{\text{DD,p}}}{\bm{\Omega}}_{\text{DD,p}}$ is diagonal, the channel estimation error is minimized \cite{werf_optimal_2026}, which is equivalent to an AF with the lowest sidelobes. However, while both metrics consistently employ the pilot structure, the interaction between data and pilot symbols still requires further optimization.
\end{remark}

\section{OTFS Signal Optimization}
\label{sec:Opt_prob}
In this section, we formulate and solve a joint OTFS DFRC optimization problem w.r.t. the pilot symbols ($\mathbf{x}_\text{DD,p}$) and the data-symbol power ($p_d$) to maximize the weighted sum of negative ISL and SINR. Both metrics account for the contributions of data power and pilot symbols as well as their interaction, thereby forming a comprehensive DFRC optimization framework that considers the channel estimation error. By optimizing $\mathbf{x}_\text{DD,p}$, we obtain the placement and power allocation of pilot symbols, while $p_d$ can reflect the power of the data payload (both pilot and data serve to offer effective communication).

This section is organized as follows. Section~\ref{sec:Opt_prob:Prob_formulation} formulates the optimization problem. Based on the formulation, we introduce an auxiliary variable to relax the original problem so that the AO method could be effectively applied to decompose the original problem into two tractable subproblems in Section~\ref{sec:Opt_prob:Alter_opt}. Subsequently, we address the first subproblem to optimize $p_d$ using convex optimization in Section~\ref{sec:Opt_prob:solve_AO_1}. Section~\ref{sec:Opt_prob:solve_AO_2} then tackles the second subproblem w.r.t. $\mathbf{x}_\text{DD,p}$ within an ADMM-SCA framework. Finally, the overall optimization framework is summarized in Section~\ref{sec:Opt_prob:sum}.

\subsection{Problem Formulation}
\label{sec:Opt_prob:Prob_formulation}
Using the defined $\text{SINR}$ in (\ref{equ:opt_SINR}) and $\text{ISL}$ in (\ref{equ:ISL_simp}), we first formulate the OTFS DFRC optimization problem as follows. In contrast to \cite{werf_optimal_2026}, we preserve full degrees of freedom for pilot symbols in joint SINR–ISL DFRC optimization.
\begin{subequations} \label{equ:P1_OBJ}
\begin{align}
    \mathop {\max }\limits_{p_d,{\mathbf{x}_\text{DD,p}}} \quad 
    & \eta \,\text{SINR}\left(p_d,{\mathbf{x}_\text{DD,p}}\right) - \left(1 - \eta\right)\text{ISL}\left(p_d,{\mathbf{x}_\text{DD,p}}\right), \label{equ:objective_sub} \\
    \text{s.t.} \quad 
    & P_{\text{T}}\left(p_d,{\mathbf{x}_\text{DD,p}}\right) \leq {P_\text{max}}, \label{equ:P1_C1} \\
    & \overline{f}_{00}\left(p_d,{\mathbf{x}_\text{DD,p}}\right) \geq \xi_{\text{min}} , \label{equ:P1_C2}
\end{align}
\end{subequations}
\begin{subequations}
\begin{align}
    \text{with }P_{\text{T}}\left(p_d,{\mathbf{x}_\text{DD,p}}\right) &\triangleq \tfrac{1}{MN+N_{\text{CP}}} \overline{f}_{00}\left(p_d,{\mathbf{x}_\text{DD,p}}\right),\\
    \overline{f}_{00}\left(p_d,{\mathbf{x}_\text{DD,p}}\right) &\triangleq \varepsilon\left\{{f}_{00}\right\}\\
    &=p_d\operatorname{Tr}\big( 
        {{\bm{\Phi}}_{\text{d}}^H {\bf{B}}^H {\bf{B}} {{\bm{\Phi}}_{\text{d}}}} 
        \big) \nonumber \\ &\quad +
        {{\mathbf{x}}_{{\text{DD,p}}}^H{\bm{\Phi}}_{\text{p}}^H {\bf{B}}^H {\bf{B}} 
        {\bm{\Phi}}_{\text{p}}{\mathbf{x}}_{{\text{DD,p}}}},
\end{align}
\end{subequations}
where $\eta$ is the ISAC trade-off weight. (\ref{equ:P1_C1}) is the transmit power constraint, where $P_\text{max}$ is the transmit power budget, and $P_{\text{T}}$ is the actual emitted power, which could be easily derived from the DD domain.
(\ref{equ:P1_C2}) imposes a constraint on the mainlobe of the AF to ensure a significant gap between the mainlobe and the sidelobe for radar applications~\cite{zhang_multi_functional_2024}, where $\mathbf{B} \triangleq {\bf{\Gamma }}\left( {{\bf{F}}_N^H \otimes {{\bf{I}}_M}} \right)$. Specifically, in (\ref{equ:P1_C2}), $\overline{f}_{00}\left(p_d,{\mathbf{x}_\text{DD,p}}\right)$ is the mainlobe of the expected AF of the transmitted signal, and $\xi_{\text{min}}$ is the minimum mainlobe requirement for the radar application.

\subsection{Alternating Optimization}
\label{sec:Opt_prob:Alter_opt}
To solve problem (\ref{equ:P1_OBJ}), we aim to reformulate it into convex subproblems. First, in problem (\ref{equ:P1_OBJ}), the communication term in the objective function, $\text{SINR}$, is non-concave and involves complex matrix inversion as shown in (\ref{equ:opt_SINR}). To address this, we introduce an auxiliary variable $s_1$ for (\ref{equ:opt_SINR}) to simplify the matrix inversion as follows
\begin{equation}
    {\text{SINR}^\prime}\left(p_d,s_1\right) = \frac{{p_d}}{{\sigma_{\mathbf{n}}^2}}{{\left[ {\frac{{p_d}}{{\sigma_{\mathbf{n}}^2}} s_1 + 1} \right]}^{ - 1}}, \label{equ:SINR_s1}
\end{equation}
after which the following constraint should be introduced to keep the equivalence
\begin{equation}
    s_1 \ge {\operatorname{Tr}}\left( {\sigma _{\mathbf{h}}^2{{\left( {{{\mathbf{I}}_{{K_h}}} + \frac{{\sigma _{\mathbf{h}}^2}}{{\sigma _{\mathbf{n}}^2}}{{\bm{\Omega}}}_{{\text{DD,p}}}^H{{{\bm{\Omega}}}_{{\text{DD,p}}}}} \right)}^{ - 1}}} \right).
\end{equation}
As a consequence, the optimization problem (\ref{equ:P1_OBJ}) is reformulated as follows
\begin{subequations} \label{equ:P2}
\begin{align}
    \mathop {\max }\limits_{p_d, s_1, {\mathbf{x}_\text{DD,p}}} \quad 
    & \eta \,{\text{SINR}^\prime}\left(p_d,s_1\right) - \left(1 - \eta\right)\text{ISL}\left(p_d,{\mathbf{x}_\text{DD,p}}\right), \label{equ:P2_objective} \\
    \text{s.t.} \quad 
    & s_1 \geq \operatorname{Tr}\left[ 
        \sigma _{\mathbf{h}}^2 \bigl( 
        \mathbf{I}_{K_h} + 
        \mathbf{\Theta}\left(\mathbf{x}_{\text{DD,p}}\right) \bigr)^{-1} \right], \label{equ:P2_C1} \\
    & (\ref{equ:P1_C1})\text{ and }(\ref{equ:P1_C2}), \label{equ:P2_C}
\end{align}
\end{subequations}
\begin{align}
    \text{with }\mathbf{\Theta}\left(\mathbf{x}_{\text{DD,p}}\right) \buildrel \Delta \over= \; & \frac{\sigma _{\mathbf{h}}^2}{\sigma _{\mathbf{n}}^2} \left(\mathbf{I}_{K_h} \otimes \mathbf{x}_{\text{DD,p}}^H\right) \widetilde{{\bm{\Omega}}}_{\text{DD,p}}^H \nonumber\\ &\cdot\widetilde{{\bm{\Omega}}}_{\text{DD,p}} \left(\mathbf{I}_{K_h} \otimes \mathbf{x}_{\text{DD,p}}\right),
\end{align}
where $\widetilde{{\bm{\Omega}}}_{\text{DD,p}}$ is the extended dictionary matrix of pilot symbols defined similarly to $\widetilde{{\bm{\Omega}}}_{\text{DD,d}}$ as in (\ref{equ:Omega_tilde_c}), with ${\bm{\Omega}}_{\text{DD,p}} = \widetilde{{\bm{\Omega}}}_{\text{DD,p}}\left(\mathbf{I}_{K_h} \otimes \mathbf{x}_\text{DD,p}\right)$.

From (\ref{equ:P2}), we observe that $p_d$ is only related to constraints (\ref{equ:P2_C}), and the objective function is concave w.r.t. $p_d$ for fixed 
$s_1$ and $\mathbf{x}_\text{DD,p}$. We therefore adopt an AO framework so as to exploit this concavity. The resultant subproblems are expressed as follows, where $n$ denotes the iteration index of the AO algorithm.

\emph{1) AO Problem 1:} This subproblem optimizes the power of the data symbols, \(p_d\), given fixed $s_{1}^{(n)}$ and ${\mathbf{x}}^{(n)}_{{\text{DD}},{\text{p}}}$, which could be formulated as follows
\begin{subequations} \label{equ:SP1_simp}
\begin{align}
\max_{p_d}\;&\eta\,\text{SINR}'\bigl(p_d,s_1^{(n)}\bigr)-\left(1 - \eta\right)\,\text{ISL}\bigl(p_d,\mathbf{x}_\text{DD,p}^{(n)}\bigr)\label{equ:SP1_simp_objective}\\
\text{s.t.}\;&P_\text{T}\bigl(p_d,\mathbf{x}_\text{DD,p}^{(n)}\bigr)\le P_\text{max},\label{equ:SP1_simp_d1}\\
&\overline{f}_{00}\bigl(p_d,\mathbf{x}_\text{DD,p}^{(n)}\bigr) \ge \xi_\text{min}.\label{equ:SP1_simp_d2}
\end{align}
\end{subequations}

\emph{2) AO Problem 2:} This subproblem optimizes the auxiliary variable \(s_{1}\) and the pilot symbols \(\mathbf{x}_{\text{DD,p}}\) given fixed \(p_d^{(n)}\), and it could be expressed as follows
\begin{subequations} \label{equ:SP2_simp}
\begin{align}
    \mathop {\max }\limits_{s_1, {{\mathbf{x}}_{{\text{DD}},{\text{p}}}}} 
    & \eta \,{\text{SINR}^\prime}\left({p_d^{(n + 1)}}, s_1\right)
    - \left(1 - \eta\right)\,\text{ISL}\left({p_d^{(n + 1)}},\mathbf{x}_{\text{DD,p}}\right), 
    \label{equ:SP2_simp_objective} \\
    \text{s.t.} \quad 
    & s_1 \geq \operatorname{Tr} \left[ 
        \sigma_{\mathbf{h}}^2 \left( 
        \mathbf{I}_{K_h} + 
        \mathbf{\Theta}\left(\mathbf{x}_{\text{DD,p}}\right) 
        \right)^{-1} 
    \right], \label{equ:SP2_simp_d1} \\
    & P_\text{T}\bigl(p_d^{(n+1)},\mathbf{x}_\text{DD,p}\bigr) \leq P_\text{max}, \label{equ:SP2_simp_d2} \\
    & \overline{f}_{00}\bigl(p_d^{(n+1)},\mathbf{x}_\text{DD,p}\bigr) \ge \xi_{\text{min}}.\label{equ:SP2_simp_d3}
\end{align}
\end{subequations}

\subsection{Solving AO Problem 1}
\label{sec:Opt_prob:solve_AO_1}
The first AO problem (\ref{equ:SP1_simp}) could be equivalently rewritten as follows
\begin{subequations} \label{equ:SP1}
\begin{align}
\max_{p_d}\;&\;\eta\,p_d
\left(p_d\,s_1^{(n)} + \sigma_{\mathbf{n}}^2\right)^{-1} -\left(1 - \eta\right)\left(p^2_d\,c_{1}+p_d\,c_{2}+c_{3}\right)
\label{equ:SP1_objective}\\
\text{s.t.}\;
&\tfrac{1}{MN+N_{\text{CP}}}\left(p_d\,c_{4}+c_{5}\right)\le P_\text{max},\label{equ:SP1_C1}\\
&p_d\,c_{4}+c_{5} \ge \xi_{\text{min}},\label{equ:SP1_C2}
\end{align}
\end{subequations}
where $c_i$, \(i=1,\ldots,5\), are the corresponding constants for a fixed $\mathbf{x}^{(n)}_\text{DD,p}$ (we omit the expression here for space reasons). It is easy to see that the first AO problem is convex w.r.t. $p_d$ and could be easily solved using a convex optimization toolbox, e.g., CVX.

\subsection{Solving AO Problem 2}
\label{sec:Opt_prob:solve_AO_2}
The second AO problem~(\ref{equ:SP2_simp}) is non-convex, particularly w.r.t. ${\mathbf{x}}_{\text{DD},\text{p}}$. To this end, we substitute \( {\mathbf{x}}_{\text{DD},\text{p}} \) in all equations with \( {\mathbf{x}}_{\text{DD},\text{p,1}} \) and \( {\mathbf{x}}_{\text{DD},\text{p,2}}\), introducing a constraint \( {\mathbf{x}}_{\text{DD},\text{p,1}} = {\mathbf{x}}_{\text{DD},\text{p,2}} \) to ensure consistency. The introduced equivalence constraint then allows an ADMM framework to formulate tractable subproblems. Specifically, the original optimization subproblem of (\ref{equ:SP2_simp}) is reformulated as follows
\begin{subequations} \label{equ:SP3}
\begin{align}
    \mathop{\min}\limits_{\substack{s_{1},{{\mathbf{x}}_{\text{DD},\text{p,1}}},\\{\mathbf{A}},{{\mathbf{x}}_{\text{DD},\text{p,2}}}}} \quad
    &\! -\eta\,\text{SINR}'\left(p_d^{(n+1)},\,s_1\right) \nonumber \\
    &\! +\left(1 - \eta\right)\,\text{ISL}'\left(p_d^{(n+1)},\,\mathbf{x}_{\text{DD,p,1}},\,\mathbf{x}_{\text{DD,p,2}}\right) \nonumber \\
    &\! + \frac{\rho}{2} \Big\| 
        {{\mathbf{x}}_{{\text{DD}},{\text{p,1}}}}
        - {{\mathbf{x}}_{{\text{DD}},{\text{p,2}}}}
        + \mathbf{d} 
    \Big\|^2_2 \nonumber \\
    &\! + \frac{\zeta}{2} \Big\|
        \mathbf{A} \cdot \mathbf{\Xi}\left(\mathbf{x}_{\text{DD},\text{p,1}},\mathbf{x}_{\text{DD},\text{p,2}}\right) - \mathbf{I}
    \Big\|^2_F, \label{equ:SP3_objective}\\
    \text{s.t.} \quad
    & s_{1} \ge \operatorname{Tr}\left(\sigma_{\mathbf{h}}^2 \mathbf{A}\right), \label{equ:SP3_C1}\\
    & P_{\text{T}}\left(p_d^{(n+1)},{\mathbf{x}_\text{DD,p,1}},{\mathbf{x}_\text{DD,p,2}}\right) \le P_\text{max}, \label{equ:SP3_C2}\\
    & \overline{f}_{00}\left(p_d^{(n+1)},{\mathbf{x}_\text{DD,p,1}},{\mathbf{x}_\text{DD,p,2}}\right) \ge \xi_\text{min}, \label{equ:SP3_C3}
\end{align}
\end{subequations}
\begin{subequations}
\begin{align}
\text{with }&\;\text{ISL}'\!\left(p_d,
                    \mathbf{x}_{\text{DD,p},1},
                    \mathbf{x}_{\text{DD,p},2}\right)\notag\\
  \triangleq &\sum_{l,k} \Bigl[
      p_d^2\!\left(a_{lk} + \lvert b_{lk}\rvert^{2}\right) 
  + p_d
        \mathbf{x}_{\text{DD,p},2}^{\!H}
        \mathbf{B}_{lk}
        \mathbf{x}_{\text{DD,p},1} \notag\\
  &+ 2p_d
        \operatorname{Re}\!\left(
          b_{lk}\,
          \mathbf{x}_{\text{DD,p},2}^{\!H}
          \mathbf{A}_{\text{p},lk}^{H}
          \mathbf{x}_{\text{DD,p},1}
        \right) \notag\\
  &+ \mathbf{x}_{\text{DD,p},1}^{\!H}
        \mathbf{A}_{\text{p},lk}
        \mathbf{x}_{\text{DD,p},2}\,
        \mathbf{x}_{\text{DD,p},2}^{\!H}
        \mathbf{A}_{\text{p},lk}^{H}
        \mathbf{x}_{\text{DD,p},1}
    \Bigr],
\label{equ:ISL_simp_ADMM}
\end{align}
\begin{align}
    \,\;\;\;\text{with }\mathbf{\Xi}\left(\mathbf{x}_{\text{DD},\text{p,1}},\mathbf{x}_{\text{DD},\text{p,2}}\right) \triangleq & \;\mathbf{I}_{K_h} + \frac{\sigma_{\mathbf{h}}^2}{\sigma_{\mathbf{n}}^2} \left(\mathbf{I}_{K_h}\otimes\mathbf{x}_{\text{DD},\text{p,2}}^H\right) \nonumber \\
    & \cdot\widetilde{\bm{\Omega}}_{\text{DD},\text{p}}^H \widetilde{\bm{\Omega}}_{\text{DD},\text{p}} \left(\mathbf{I}_{K_h} 
    \otimes\mathbf{x}_{\text{DD},\text{p,1}}\right),
\end{align}
\begin{align}
    \quad\;\text{with } & P_{\text{T}}\left(p_d,{\mathbf{x}_\text{DD,p,1}},{\mathbf{x}_\text{DD,p,2}}\right) \nonumber \\ 
    \triangleq \; & \tfrac{1}{MN+N_{\text{CP}}} \overline{f}_{00}\left(p_d,{\mathbf{x}_\text{DD,p,1}},{\mathbf{x}_\text{DD,p,2}}\right),\\
    \quad\;\text{with } & \overline{f}_{00}\left(p_d,{\mathbf{x}_\text{DD,p,1}},{\mathbf{x}_\text{DD,p,2}}\right) \nonumber \\ 
    \triangleq \, & p_d\operatorname{Tr}\big( 
        {{\bm{\Phi}}_{\text{d}}^H {\bf{B}}^H {\bf{B}} {{\bm{\Phi}}_{\text{d}}}} 
        \big) +
        {{\mathbf{x}}_{{\text{DD,p,2}}}^H{\bm{\Phi}}_{\text{p}}^H {\bf{B}}^H {\bf{B}} 
        {\bm{\Phi}}_{\text{p}}{\mathbf{x}}_{{\text{DD,p,1}}}},
\end{align}
\end{subequations}
where a slack matrix variable $\mathbf{A}$ is introduced to further relax the constraint (\ref{equ:SP2_simp_d1}) as (\ref{equ:SP3_C1}), thereby avoiding matrix inversion and offering additional degrees of freedom during the optimization process. $\rho$ and $\zeta$ are penalty coefficients for the ADMM framework and the slackness operation, respectively. After introducing \( \mathbf{x}_{\text{DD,p,1}} \) and \( \mathbf{x}_{\text{DD,p,2}} \), the ISL expression is updated to~(\ref{equ:ISL_simp_ADMM}), which is separately convex w.r.t. $\mathbf{x}_\text{DD,p,1}$ and $\mathbf{x}_\text{DD,p,2}$ after the ADMM reformulation.

Through the ADMM structure, we could iteratively update the $s_1$, ${{\mathbf{x}}_{{\text{DD}},{\text{p,1}}}}$, $ {{\mathbf{x}}_{{\text{DD}},{\text{p,2}}}}$, and $\mathbf{A}$ until the solutions of ${{\mathbf{x}}_{{\text{DD}},{\text{p,1}}}}$ and ${{\mathbf{x}}_{{\text{DD}},{\text{p,2}}}}$ converge. Specifically, at the $n$-th AO iteration and $m$-th ADMM iteration, problem (\ref{equ:SP3}) is iteratively updated as follows
\begin{align}
\begin{bmatrix}
    s^{(n,m+1)}_{1}\\
    \mathbf{x}^{(n,m+1)}_{\text{DD,p,1}}
\end{bmatrix}
:=& \arg\min_{\substack{
            s_{1}, \\ 
            {{\mathbf{x}}_{{\text{DD}},{\text{p,1}}}}
        }}
\biggl\{
    -\eta\,
    \text{SINR}'\left(p_d^{(n+1)}, s_1\right) \nonumber\\
&+\left(1 - \eta\right)\,
    \text{ISL}'\left(p_d^{(n+1)}, {{\mathbf{x}}_{{\text{DD}},{\text{p,1}}}}, {{\mathbf{x}}^{(n,m)}_{{\text{DD}},{\text{p,2}}}}\right)\, \nonumber \\ 
&+ \frac{\rho}{2} \Big\| 
        {{\mathbf{x}}_{{\text{DD}},{\text{p,1}}}} 
        - {{\mathbf{x}}^{(n,m)}_{{\text{DD}},{\text{p,2}}}} 
        + \mathbf{d}^{(n,m)} 
    \Big\|^2_2 \nonumber \\
&+ \frac{\zeta}{2} \Big\|
    \mathbf{A}^{(n,m)} \mathbf{\Xi}\left(\mathbf{x}_{\text{DD},\text{p,1}},\mathbf{x}^{(n,m)}_{\text{DD},\text{p,2}}\right) - \mathbf{I}
\Big\|^2_F\nonumber \\
&+\mathbbm{1}_{\left(s_{1} \ge \operatorname{Tr}\left(\sigma_{\mathbf{h}}^2 \mathbf{A}^{(n,m)}\right)\right)} \nonumber\\
&+\mathbbm{1}_{\left(P_{\text{T}}\left(p_d^{(n+1)},{{\mathbf{x}}_{{\text{DD}},{\text{p,1}}}},{{\mathbf{x}}^{(n,m)}_{{\text{DD}},{\text{p,2}}}}\right) \le P_{\text{max}} \right)} \nonumber \\
&+\mathbbm{1}_{\left(\overline{f}_{00}\left(p_d^{(n+1)},\mathbf{x}_{\text{DD,p,1}},\mathbf{x}^{(n,m)}_{\text{DD,p,2}}\right) \ge \xi_{\text{min}} \right)}\biggr\}, \label{equ:update_x1}\\
\mathbf{x}^{(n,m+1)}_{\text{DD,p,2}}
:=& \nonumber \\ 
\arg\min_{\substack{
            {{\mathbf{x}}_{{\text{DD}},{\text{p,2}}}}
        }}
\biggl\{&\;\left(1 - \eta\right)\,
    \text{ISL}'\left(p_d^{(n+1)}, {{\mathbf{x}}^{(n,m+1)}_{{\text{DD}},{\text{p,1}}}}, {{\mathbf{x}}_{{\text{DD}},{\text{p,2}}}}\right)
\, \nonumber \\ 
&+ \frac{\rho}{2} \Big\| 
        {{\mathbf{x}}^{(n,m+1)}_{{\text{DD}},{\text{p,1}}}} 
        - {{\mathbf{x}}_{{\text{DD}},{\text{p,2}}}} 
        + \mathbf{d}^{(n,m)} 
    \Big\|^2_2 \nonumber \\
&+ \frac{\zeta}{2} \Big\|
    \mathbf{A}^{(n,m)} \mathbf{\Xi}\left(\mathbf{x}^{(n,m+1)}_{\text{DD},\text{p,1}},\mathbf{x}_{\text{DD},\text{p,2}}\right) - \mathbf{I}
\Big\|^2_F\nonumber \\
&+\mathbbm{1}_{\left(P_{\text{T}}\left(p_d^{(n+1)},{{\mathbf{x}}^{(n,m+1)}_{{\text{DD}},{\text{p,1}}}},{{\mathbf{x}}_{{\text{DD}},{\text{p,2}}}}\right) \le P_{\text{max}} \right)} \nonumber \\
&+\mathbbm{1}_{\left(\overline{f}_{00}\left(p_d^{(n+1)},\mathbf{x}^{(n,m+1)}_{\text{DD,p,1}},\mathbf{x}_{\text{DD,p,2}}\right) \ge \xi_{\text{min}} \right)}\biggr\}\,,\label{equ:update_x2}\\
\mathbf{A}^{(n,m+1)}
:=& \arg\min_{\substack{
            {\mathbf{A}}
        }}
\biggl\{ \frac{\zeta}{2} \Big\|
    \mathbf{A} \cdot \mathbf{\Xi}^{(n,m+1)} - \mathbf{I}
\Big\|^2_F\nonumber \\
&+\mathbbm{1}_{\left(s^{(n,m+1)}_{1} \ge \operatorname{Tr}\left(\sigma_{\mathbf{h}}^2 \mathbf{A}\right)\right)}\biggr\},\label{equ:update_A}\\
\mathbf{d}^{(n,m+1)}
:=& \mathbf{d}^{(n,m)} + \mathbf{x}^{(n,m+1)}_{\text{DD,p,1}} - \mathbf{x}^{(n,m+1)}_{\text{DD,p,2}},\label{equ:update_d}
\end{align}
where $\mathbf{\Xi}^{(n,m+1)} \triangleq \mathbf{\Xi}\left(\mathbf{x}^{(n,m+1)}_{\text{DD},\text{p,1}},\mathbf{x}^{(n,m+1)}_{\text{DD},\text{p,2}}\right)$ in (\ref{equ:update_A}).

\emph{1) Updating $\mathbf{x}_\text{DD,p,1}$ in (\ref{equ:update_x1}):}

In (\ref{equ:update_x1}), the SINR in the objective function is still non-convex w.r.t. $s_1$. Hence, we use SCA to linearly approximate the SINR in (\ref{equ:SINR_s1}) via a first-order Taylor expansion. Moreover, we integrate the SCA operation into the ADMM framework within the same loop. At the $n$-th iteration of the AO algorithm and the $m$-th iteration of the ADMM-SCA operation, the SCA formulation of the SINR in problem (\ref{equ:update_x1}) is given by
\begin{equation}
\begin{aligned}
&\text{SINR}_{\text{SCA}}'\left(p_d^{(n+1)},\,s^{(n,m)}_{1},\,s_1\right)\\
\buildrel \Delta \over=\;&\frac{p_d^{(n+1)}}{\sigma_{\mathbf n}^2}
\Bigl(\frac{p_d^{(n+1)}}{\sigma_{\mathbf n}^2}\,s^{(n,m)}_{1}+1\Bigr)^{-1}\\
&-\Bigl(\frac{p_d^{(n+1)}}{\sigma_{\mathbf n}^2}\Bigr)^{2}
\Bigl(\frac{p_d^{(n+1)}}{\sigma_{\mathbf n}^2}\,s^{(n,m)}_{1}+1\Bigr)^{-2}\bigl(s_1 - s^{(n,m)}_{1}\bigr)\,,
\end{aligned}
\end{equation}
where \(s^{(n,m)}_{1}\) is the solution from the previous SCA iteration and serves as the SCA operating point for the current iteration. Consequently, the optimization problem in (\ref{equ:update_x1}) becomes convex as follows

\begin{subequations} \label{equ:ADMM-x1}
\begin{align}
    \mathop {\min }\limits_{
        \substack{
            s_{1}, \\ 
            {{\mathbf{x}}_{{\text{DD}},{\text{p,1}}}}
        }
    } \quad 
    & -\eta\,
    \text{SINR}_{\text{SCA}}'\left(p_d^{(n+1)},\,s^{(n,m)}_{1},\,s_1\right) \nonumber\\
    &+\left(1 - \eta\right)\,
        \text{ISL}'\left(p_d^{(n+1)}, {{\mathbf{x}}_{{\text{DD}},{\text{p,1}}}}, {{\mathbf{x}}^{(n,m)}_{{\text{DD}},{\text{p,2}}}}\right)\,
    \nonumber \\
    & + \frac{\rho}{2} \Big\| 
        {{\mathbf{x}}_{{\text{DD}},{\text{p,1}}}} 
        - {{\mathbf{x}}^{(n,m)}_{{\text{DD}},{\text{p,2}}}} 
        + \mathbf{d}^{(n,m)}
    \Big\|^2_2 \nonumber \\
    & + \frac{\zeta}{2} \Big\| 
        \mathbf{A}^{(n,m)} \cdot \mathbf{\Xi}\left(\mathbf{x}_{\text{DD},\text{p,1}},\mathbf{x}^{(n,m)}_{\text{DD},\text{p,2}}\right) - \mathbf{I}
    \Big\|^2_F, \label{equ:ADMM-x1-objective} \\
    \text{s.t.} \quad 
    & s_{1} \geq \operatorname{Tr}\left( 
        \sigma _{\mathbf{h}}^2\mathbf{A}^{(n,m)} 
    \right), \label{equ:ADMM-x1-C1} \\
    & P_{\text{T}}\left(p_d^{(n+1)},{\mathbf{x}_\text{DD,p,1}},{\mathbf{x}_\text{DD,p,2}^{(n,m)}}\right) \leq P_\text{max}, \label{equ:ADMM-x1-C2} \\
    & \overline{f}_{00}\left(p_d^{(n+1)},{\mathbf{x}_\text{DD,p,1}},{\mathbf{x}_\text{DD,p,2}^{(n,m)}}\right) \geq \xi_{\text{min}}, \label{equ:ADMM-x1-C3}
\end{align}
\end{subequations}
It could be observed that $\text{SINR}_{\text{SCA}}'(p_d^{(n+1)},s^{(n,m)}_{1},s_1)$ is affine w.r.t. $s_1$, while $\text{ISL}'(p_d^{(n+1)},\mathbf{x}_{\text{DD,p,1}},\mathbf{x}^{(n,m)}_{\text{DD,p,2}})$ and the penalty terms are convex w.r.t. $\mathbf{x}_{\text{DD,p,1}}$ in the objective function. Therefore, the objective in \eqref{equ:ADMM-x1} is convex w.r.t. $s_1$ and $\mathbf{x}_{\text{DD,p,1}}$. Since the constraints are affine, the problem \eqref{equ:ADMM-x1} is convex and hence could be easily solved.

\emph{2) Updating ${{\mathbf{x}}_{{\text{DD}},{\text{p,2}}}}$ in (\ref{equ:update_x2}):}

Problem (\ref{equ:update_x2}) could be reformulated as follows
\begin{subequations} \label{equ:ADMM-x2}
\begin{align}
    \mathop {\min }\limits_{
        {{\mathbf{x}}_{{\text{DD}},{\text{p,2}}}}
    } \quad 
    & \;\left(1 - \eta\right)\,
    \text{ISL}'\left(p_d^{(n+1)}, {{\mathbf{x}}^{(n,m+1)}_{{\text{DD}},{\text{p,1}}}}, {{\mathbf{x}}_{{\text{DD}},{\text{p,2}}}}\right)
    \, \nonumber \\
    & + \frac{\rho}{2} \Big\| 
        {{\mathbf{x}}^{(n,m + 1)}_{{\text{DD}},{\text{p,1}}}} 
        - {{\mathbf{x}}_{{\text{DD}},{\text{p,2}}}} 
        + \mathbf{d}^{(n,m)} 
    \Big\|^2_2 \nonumber \\
    & + \frac{\zeta}{2} \Big\| 
        \mathbf{A}^{(n,m)} \cdot \mathbf{\Xi}\left(\mathbf{x}^{(n,m+1)}_{\text{DD},\text{p,1}},\mathbf{x}_{\text{DD},\text{p,2}}\right) - \mathbf{I}
    \Big\|^2_F, \label{equ:ADMM-x2-objective} \\
    \text{s.t.} \quad 
    & P_{\text{T}}\left(p_d^{(n+1)},{\mathbf{x}_\text{DD,p,1}^{(n,m+1)}},{\mathbf{x}_\text{DD,p,2}}\right) \leq P_\text{max}, \label{equ:ADMM-x2-C1} \\
    & \overline{f}_{00}\left(p_d^{(n+1)},{\mathbf{x}_\text{DD,p,1}^{(n,m+1)}},{\mathbf{x}_\text{DD,p,2}}\right) \geq \xi_{\text{min}}. \label{equ:ADMM-x2-C2}
\end{align}
\end{subequations}
Similar to problem~(\ref{equ:ADMM-x1}), problem~(\ref{equ:ADMM-x2}) is convex given fixed $p_d^{(n+1)}$, $s_1^{(n+1)}$, $\mathbf{x}_{\text{DD,p,1}}^{(n,m+1)}$, and $\mathbf{A}^{(n,m)}$: its objective is convex and its constraints are affine, so it could be solved directly.

\emph{3) Updating $\mathbf{A}$ in (\ref{equ:update_A}):}
\begin{subequations} \label{equ:ADMM-A}
\begin{align}
    \mathop {\min }\limits_{
        {\mathbf{A}}
    } \quad 
    & \frac{\zeta}{2} \Big\| 
        \mathbf{A} \cdot \mathbf{\Xi}^{(n,m+1)} - \mathbf{I} 
    \Big\|^2_F, \label{equ:ADMM-A-objective} \\
    \text{s.t.} \quad 
    & s^{(n,m + 1)}_{1} \geq \operatorname{Tr}\left( 
        \sigma _{\mathbf{h}}^2\mathbf{A} 
    \right). \label{equ:ADMM-A-C1} 
\end{align}
\end{subequations}
Given that $s^{(n,m + 1)}_{1}$ and $\mathbf{\Xi}^{(n, m+1)}$ are fixed, the update problem (\ref{equ:ADMM-A}) reduces to a quadratic program with a single affine constraint, and thus could be solved by a standard convex optimization solver.

\emph{4) Updating $\mathbf{d}$ in (\ref{equ:update_d}):}

The step vector ${\mathbf{d}}^{(n,m)}$ for the ADMM structure is updated as follows
\begin{align}
    {\mathbf{d}}^{(n,m+1)} = {\mathbf{d}}^{(n,m)} + {{{\mathbf{x}}^{(n,m+1)}_{{\text{DD}},{\text{p,1}}}} - {{\mathbf{x}}^{(n,m+1)}_{{\text{DD}},{\text{p,2}}}}}.
\end{align}

\begin{algorithm}
\caption{ADMM-SCA Optimization for $s_1$ and $\mathbf{x}_{\text{DD,p}}$}
\label{alg:SP3}
\begin{algorithmic}[1]
    \STATE \textbf{Input:} $\eta,\;\rho,\;\zeta,\;p_d^{(n+1)},\;\mathbf{x}^{(n)}_{\text{DD,p}},\;P_\text{max},\;\xi_\text{min},\;\epsilon_{1}>0$
    \STATE \textbf{Initialize:} 
    $m\gets0,\;s_1^{(n,0)}\gets s_1^{(n)}$
    \STATE \quad $\mathbf{x}_{\text{DD,p,1}}^{(n,0)}\gets\mathbf{x}^{(n)}_{\text{DD,p}},\;
                 \mathbf{x}_{\text{DD,p,2}}^{(n,0)}\gets\mathbf{x}^{(n)}_{\text{DD,p}},\;
                 \mathbf{d}^{(n,0)}\gets\mathbf{0}$
    \STATE \textbf{Repeat} (\textbf{Solving subproblems in} (\ref{equ:ADMM-x1})--(\ref{equ:ADMM-A})):
        \STATE \quad Solve (\ref{equ:ADMM-x1}) for $s_1^{(n,m+1)}$ and $\mathbf{x}_{\text{DD,p,1}}^{(n,m+1)}$
        \STATE \quad Solve (\ref{equ:ADMM-x2}) for $\mathbf{x}_{\text{DD,p,2}}^{(n,m+1)}$
        \STATE \quad Solve (\ref{equ:ADMM-A}) for $\mathbf{A}^{(n,m+1)}$
        \STATE \quad Update 
        $\mathbf{d}^{(n,m+1)}
            =\mathbf{d}^{(n,m)}
             +\mathbf{x}_{\text{DD,p,1}}^{(n,m+1)}
                        -\mathbf{x}_{\text{DD,p,2}}^{(n,m+1)}$
        \STATE \quad $m\gets m+1$
    \STATE \quad Quit if $\|\mathbf{x}_{\text{DD,p,1}}^{(n,m)} - \mathbf{x}_{\text{DD,p,2}}^{(n,m)}\|_2 \le \epsilon_{1}\|\mathbf{x}_{\text{DD,p,1}}^{(n,m)}\|_2$
    \STATE \textbf{Output:}
    \STATE \quad $s_1^{(n+1)}\gets s_1^{(n,m)}$
    \STATE \quad $\mathbf{x}^{(n+1)}_{\text{DD,p}}\gets\bigl(\mathbf{x}_{\text{DD,p,1}}^{(n,m)}+\mathbf{x}_{\text{DD,p,2}}^{(n,m)}\bigr)/2$
\end{algorithmic}
\end{algorithm}

The ADMM-SCA algorithm for solving the reformulated AO problem 2, i.e., problem (\ref{equ:SP3}), is summarized in Algorithm~\ref{alg:SP3}.

\subsection{Overall Optimization Structure}
\label{sec:Opt_prob:sum}
Based on the previous analysis, the overall optimization structure -- incorporating the AO framework, ADMM structure, and SCA operation for optimizing $p_d^{(n)}, s_{1}^{(n)}, \mathbf{x}^{(n)}_{\text{DD,p}}$ -- is summarized as Algorithm \ref{alg:overall}.

\begin{algorithm}
\caption{AO Structure for Problem (\ref{equ:P2})}
\label{alg:overall}
\begin{algorithmic}[1]
    \STATE \textbf{Input:} $\eta,\;\rho,\;\zeta,\;P_\text{max},\;\xi_\text{min},\epsilon_2>0$
    \STATE \textbf{Initialize:} 
    $n\gets0,\;p_d^{(n)},\;s_{1}^{(n)},\;\mathbf{x}^{(n)}_{\text{DD,p}}$
    \STATE \textbf{Repeat} (\textbf{Solving problems in (\ref{equ:SP1}) and (\ref{equ:SP3})})
        \STATE \quad Update 
            $p_d^{(n+1)}$ by solving (\ref{equ:SP1})
        \STATE \quad Update 
            $s_{1}^{(n+1)}$ and $\mathbf{x}_{\text{DD,p}}^{(n+1)}$ via Algorithm~\ref{alg:SP3}
        \STATE \quad $n\gets n+1$
        \STATE \quad Quit if $\left|\text{Obj}\left(n + 1\right) - \text{Obj}\left(n\right)\right| \le \epsilon_{2}\left|\text{Obj}\left(n + 1\right)\right|$
    \STATE \textbf{Output:} 
    $p_d^{(n+1)},\;s_{1}^{(n+1)},\;\mathbf{x}^{(n+1)}_{\text{DD,p}}$
\end{algorithmic}
\end{algorithm} Here, we define
\begin{equation}
    \text{Obj}(n) \triangleq \eta\,\text{SINR}'\bigl(p_d^{(n)},s_{1}^{(n)}\bigr)
     - \left(1 - \eta\right)\,\text{ISL}\bigl(p_d^{(n)},\mathbf{x}^{(n)}_{\text{DD,p}}\bigr).
\end{equation}

In summary, the overall optimization framework consists of a two-level loop based on an AO structure. The ADMM structure is employed in the second AO subproblem to convexify the optimization.

\section{Numerical Results}
\label{sec:Num_res}
In this section, we investigate the simulation results of the previous analysis and the OTFS DFRC signal obtained from the optimization method. In Section~\ref{sec:Num_res:cclb}, we first present the channel capacity lower bound under different channel mobility environments to gain a first insight into OTFS's superiority in handling high velocities. Subsequently, in Section~\ref{sec:Num_res:Opt_res}, we verify the OTFS DFRC results and performance obtained from the proposed optimization algorithm, and we also verify the effectiveness of our derived sensing (empirical ISL) and communication (capacity lower bound) metrics by AF and throughput performance, respectively.

In all simulations, the carrier frequency is set to 3.5 GHz, the subcarrier spacing to 15 kHz, and the transmit power to 30 dBm. The channel gains are modeled as zero-mean CSCG random variables with a variance of 0.5. The OTFS DD-domain symbol arrangement is the same as shown in Fig.~\ref{fig:OTFS_mod}, where the pilot and data symbols are separated by a GI.
% \begin{figure}[h]
%     \centering
%     \includegraphics[width=0.9\linewidth]{img/Scheme_illu.png}
%     \caption{The applied OTFS symbol arrangement in the DD domain.}
%     \label{fig:OTFS_data_arrange}
% \end{figure}

\subsection{Channel Capacity Lower Bound Justification in OTFS}
In this subsection, we demonstrate that our derived capacity lower bound accurately reflects the performance of OTFS and OFDM in various scenarios. This allows us to gain a deeper understanding of the advantages of OTFS over OFDM, especially in high-mobility situations. The channel capacity lower bounds based on~(\ref{equ:OTFS_channel_capacity_lower_bound_mat}) are evaluated via Monte Carlo simulations by assigning random Gaussian symbols to both the pilot and data regions. Unless otherwise specified, simulations of channel capacity lower bound are conducted over a high-mobility multipath channel to compare the performance of various OTFS implementations with that of conventional OFDM. The OFDM employs the same number of subcarriers as OTFS but adopts one-dimensional modulation and uniform power allocation.

\begin{table}[ht]
    \centering
    \caption{Simulation parameters for capacity lower bound}
    \begin{tabularx}{\linewidth}{X|c}
        \hline
        \textbf{Parameter} & \textbf{Value} \\ \hline \hline
        Number of subcarriers $M$ & 4 \\ \hline
        OFDM slots within one OTFS frame $N$ & 64 \\ \hline
        Pilot ratio $r_\text{pilot}$ & 0.25 \\ \hline
        CP ratio $r_\text{CP}$ & 0.75 \\ \hline
        Number of channel realizations & 2500 \\ \hline
    \end{tabularx}
    \label{tab:sim_LB_param}
\end{table}
The OTFS symbol setups for channel capacity lower bound simulation are summarized in Table II. The pilot ratio represents the proportion of pilot symbols to the total number of transmit symbols in the DD domain for OTFS, denoted as $r_\text{pilot} \triangleq \frac{K_p}{K_p+K_d}$. For OFDM, the pilot ratio is defined as the ratio of pilot time slots to the total number of time slots; additionally, $r_\text{GI}$ is the proportion of the GI to the total number of grids in the DD domain, denoted as $r_\text{GI} \triangleq \frac{MN-K_p-K_d}{MN}$. The CP ratio indicates the proportion of the CP length relative to the original OTFS frame, i.e., $r_\text{CP} = \frac{N_\text{CP}}{MN}$ for OTFS, or the original OFDM slot, i.e., $r_\text{CP} = \frac{N_\text{CP}}{M}$ for OFDM. Similar to the OTFS case, the OFDM capacity lower bound is derived by setting $N=1$ and periodically inserting pilot symbol slots among data symbol slots, while assuming that all slots experience identical channel conditions.

% \begin{figure}
%     \centering
%     \includegraphics[width=\linewidth]{img/Capacity_mp_static.pdf}
%     \caption{The channel capacity lower bound performance for a multipath static channel at the maximum channel delay and Doppler ($L = 3, Q = 0$).}
%     \label{fig:channel capacity lower bound_Dop0_Delay3}
% \end{figure}
\begin{figure}
    \centering
    \includegraphics[width=0.91\linewidth]{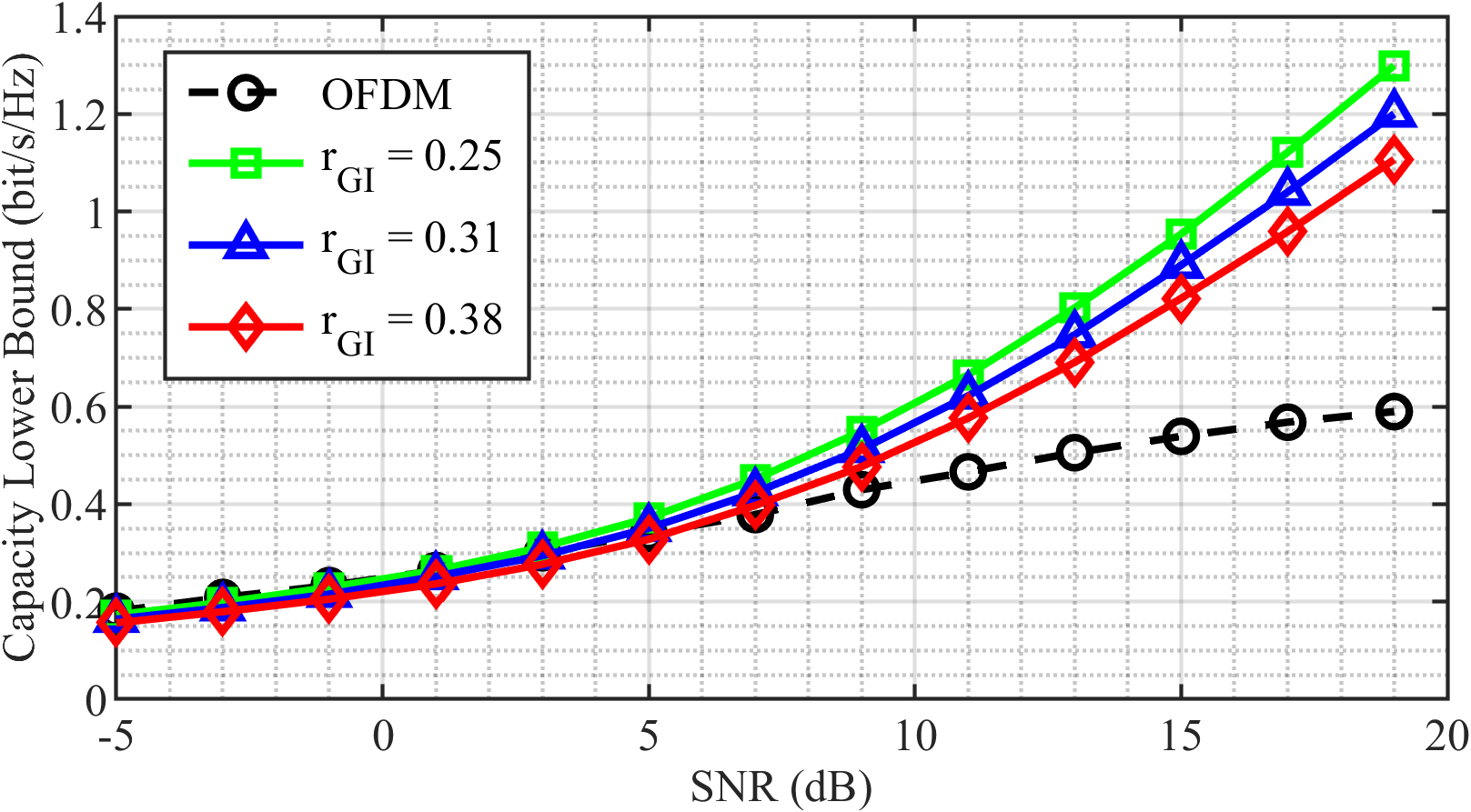}
    \caption{The channel capacity lower bound performance for a doubly-selective channel at the maximum channel delay and Doppler ($L = 3, Q = 8$).}
    \label{fig:channel capacity lower bound_Dop15}
\end{figure}
\begin{figure}
    \centering
    \includegraphics[width=0.91\linewidth]{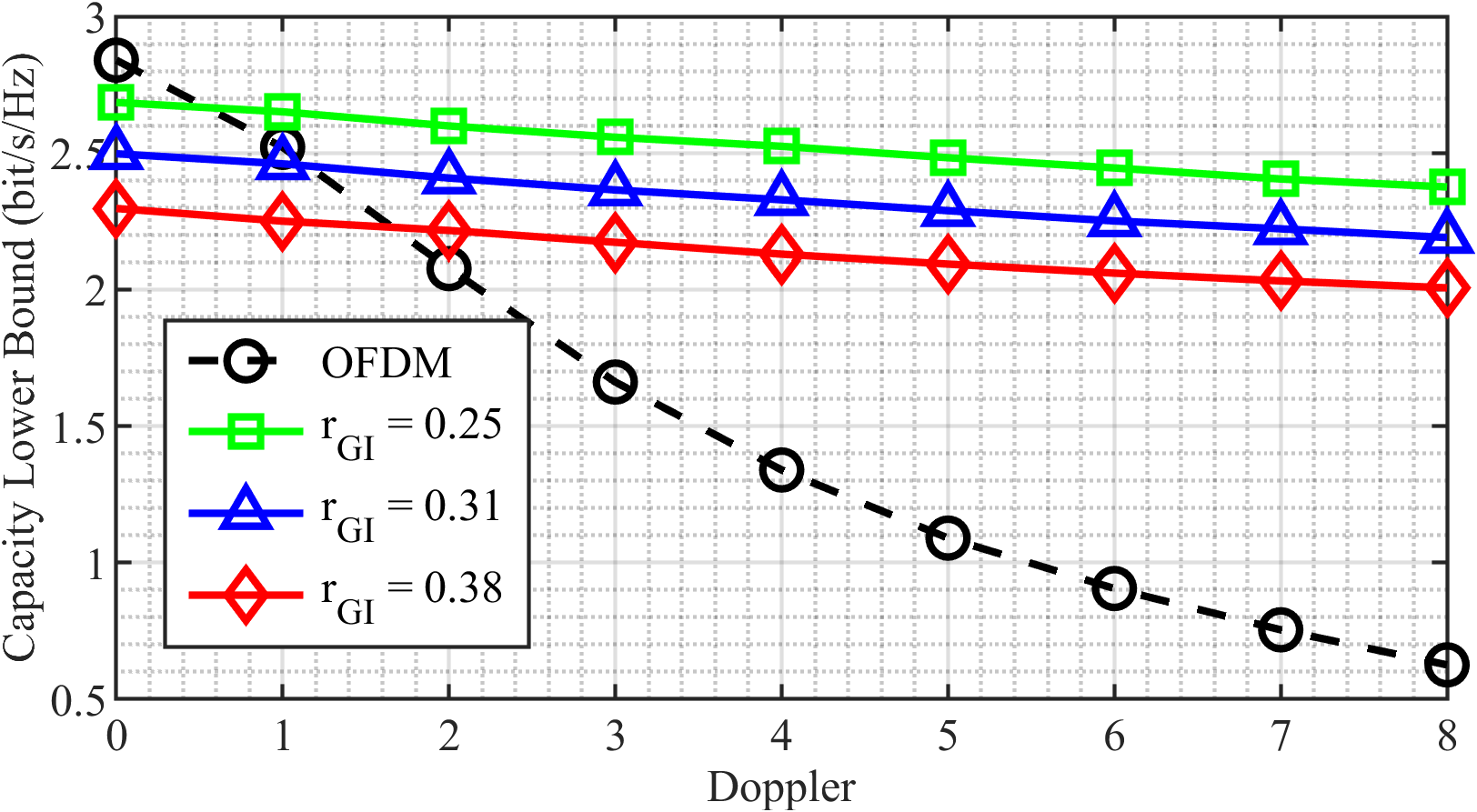}
    \caption{The channel capacity lower bound performance for OTFS and OFDM at SNR = 30 dB versus velocity, and at the maximum channel delay ($L = 3$).}
    \label{fig:channel capacity lower bound_SNR30dB}
\end{figure}

\label{sec:Num_res:cclb}
As shown in Fig.~\ref{fig:channel capacity lower bound_Dop15}, OTFS maintains a monotonically increasing capacity lower bound under doubly-selective channels, whereas OFDM degrades and saturates in the high signal-to-noise ratio (SNR) regime. The underlying reason is that OTFS employs two-dimensional modulation and DD-domain estimation that jointly resolve and track delay and Doppler. By contrast, OFDM models the channel as static, with channel dynamics appearing as residual ICI, leading to estimation error. Consistent with this mechanism, Fig.~\ref{fig:channel capacity lower bound_SNR30dB} plots the capacity lower bound versus channel mobility and shows that OTFS degrades far less than OFDM as mobility increases, confirming OTFS's superior resilience to doubly selective fading via DD-domain estimation and 2D equalization, thereby avoiding high-SNR saturation.

\subsection{DFRC Performance Verification for OTFS}
\label{sec:Num_res:Opt_res}
\begin{table}[h]
    \centering
    \caption{Parameters for OTFS DFRC optimization}
    \begin{tabularx}{\linewidth}{X|c}
        \hline
        \textbf{Parameter} & \textbf{Value} \\ \hline \hline
        Number of subcarriers $M$ & 8 \\ \hline
        OFDM slots within one OTFS frame $N$ & 16 \\ \hline
        Number of data symbols in the DD domain $K_d$ & 40 \\ \hline
        Number of pilot symbols in the DD domain $K_p$ & 24 \\ \hline
        CP ratio $r_\text{CP}$ & 0.125 \\ \hline
        Transmission SNR & 0 dB \\ \hline
        Maximum delay of the channel $L$ & 7 \\ \hline
        Maximum Doppler shift of the channel $Q$ & 3 \\ \hline  
        Number of channel coefficients $K_h$ & 32 \\ \hline  
        Maximum delay considered for ISL $\widehat{L}$ & 7 \\ \hline
        Maximum Doppler considered for ISL $\widehat{Q}$ & 5 \\ \hline
    \end{tabularx}
    \label{tab:opt-param}
\end{table}

\begin{figure}[h]   
  \centering
  \subfloat[Spike]{%      
    \includegraphics[width=0.31\linewidth]{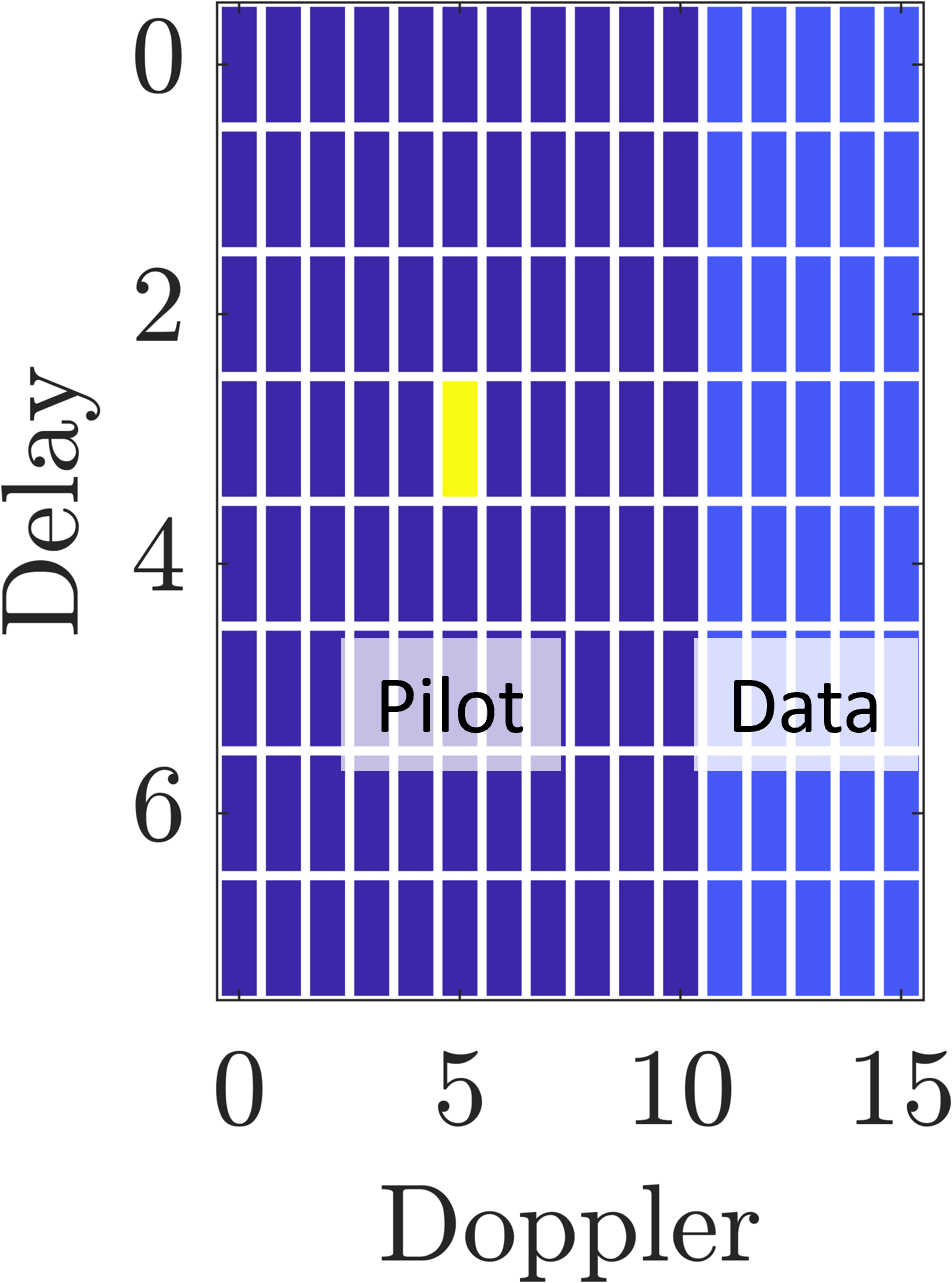}%
    \label{fig:Scheme_spike}%
  }\hfill
  \subfloat[Flat]{%
    \includegraphics[width=0.31\linewidth]{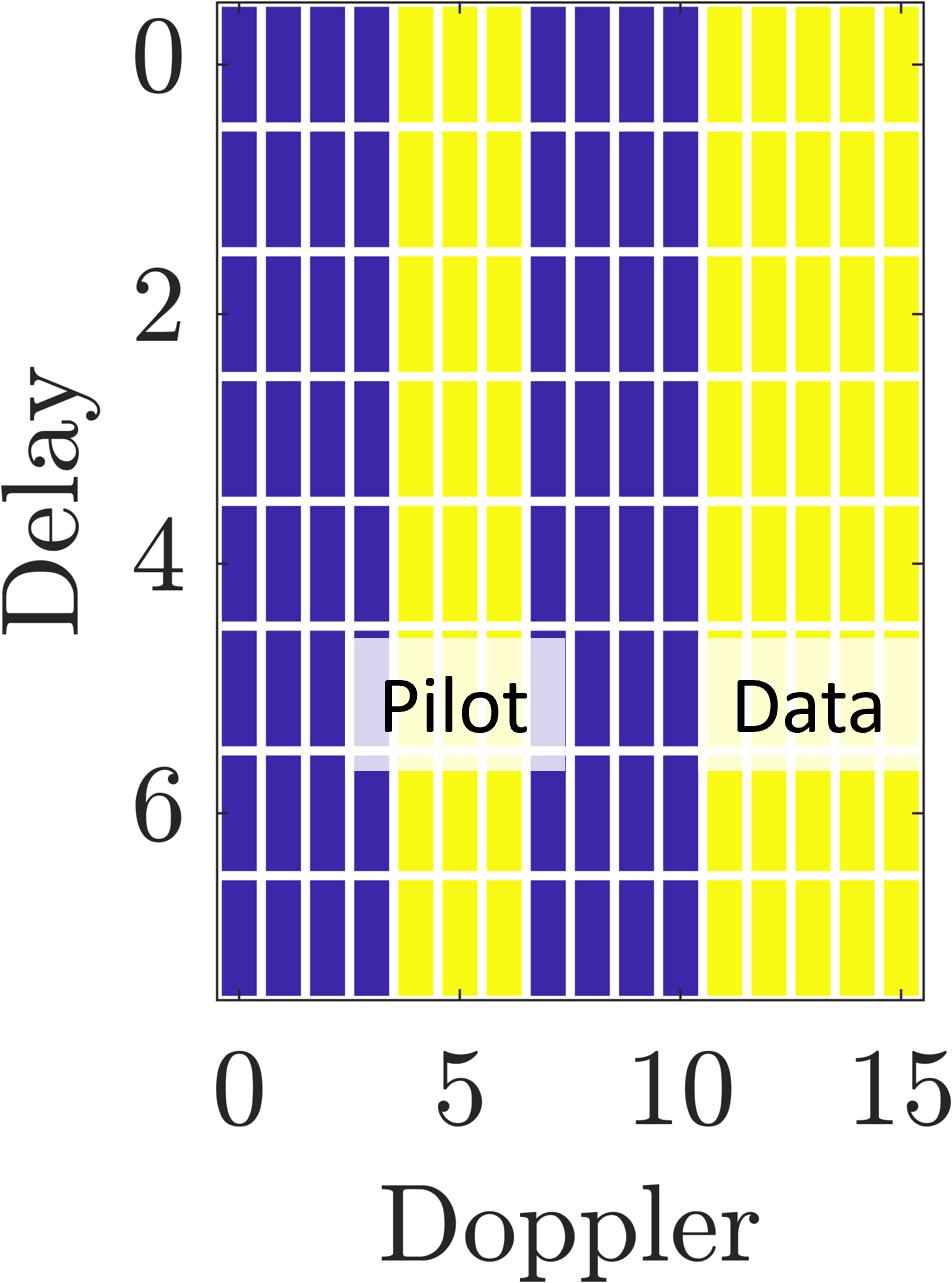}%
    \label{fig:Scheme_flat}%
  }\hfill
  \subfloat[Cluster]{%
    \includegraphics[width=0.31\linewidth]{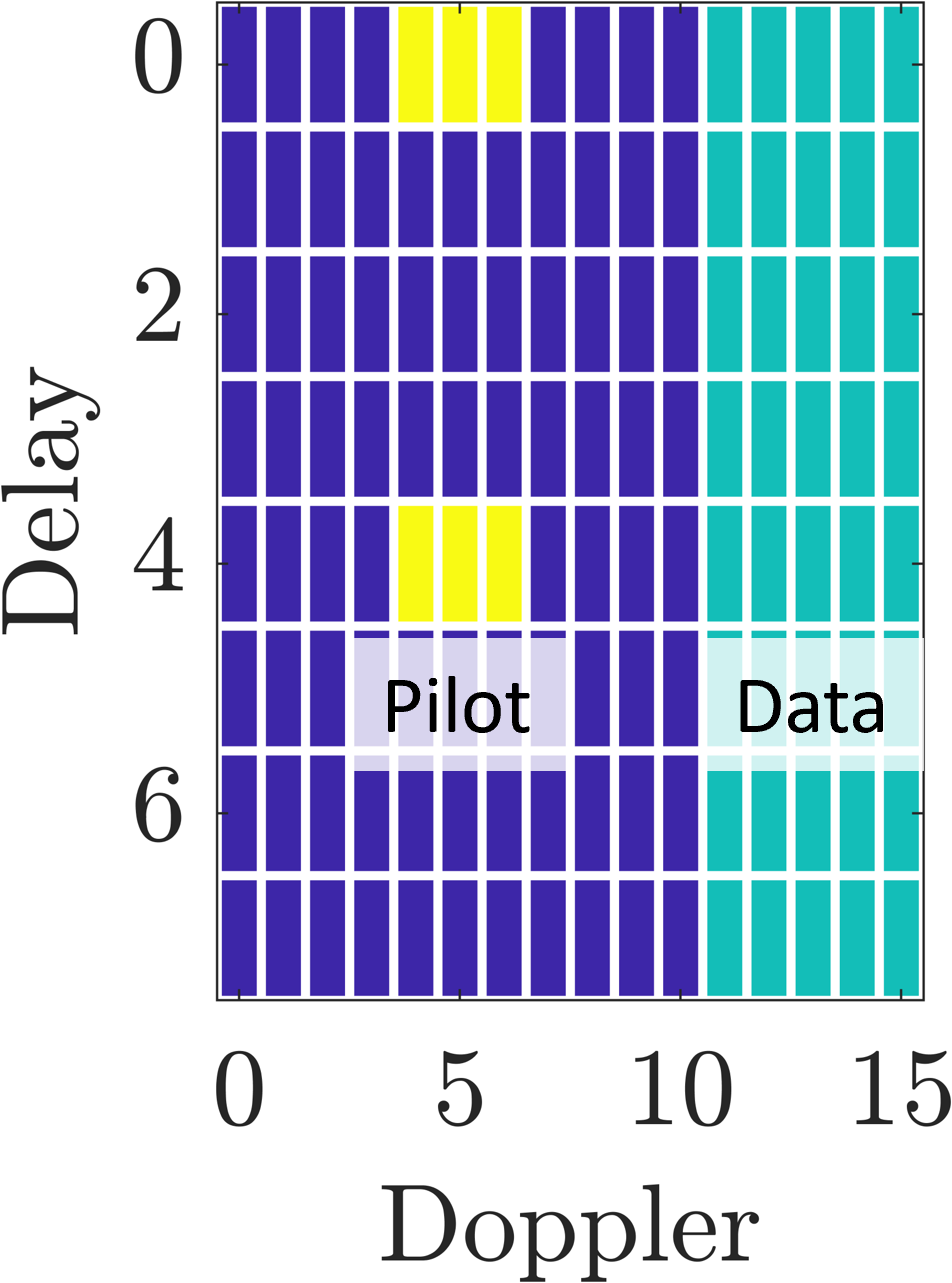}%
    \label{fig:Scheme_cluster}%
  }
  \caption{Spike, flat, and cluster arrangements in the DD domain.}
  \label{fig:Scheme}
\end{figure}

This subsection evaluates the performance of our proposed signal for OTFS DFRC. The parameters of the optimization are summarized in Table~\ref{tab:opt-param}. To reduce the chance of getting stuck in a local optimum, the optimization has been tested using several starting patterns\footnote{The optimization is conducted by normalizing the communication and sensing metrics using their respective resultant values at $\eta = 1$.}: spike, flat, and cluster arrangements, as shown in Fig.~\ref{fig:Scheme}. We also experimented with various power splits between the data and pilot regions for initialization. The experiments show that, regardless of the initial pattern, the optimized solution consistently converges to a peak-like arrangement. Different peak locations may occur, but they are equivalent up to a circular shift in the DD domain. Therefore, we adopt the spike arrangement as the initialization in the subsequent optimizations, since it leads to faster convergence.

\begin{figure}[h]
    \centering
    \includegraphics[width=0.9\linewidth]{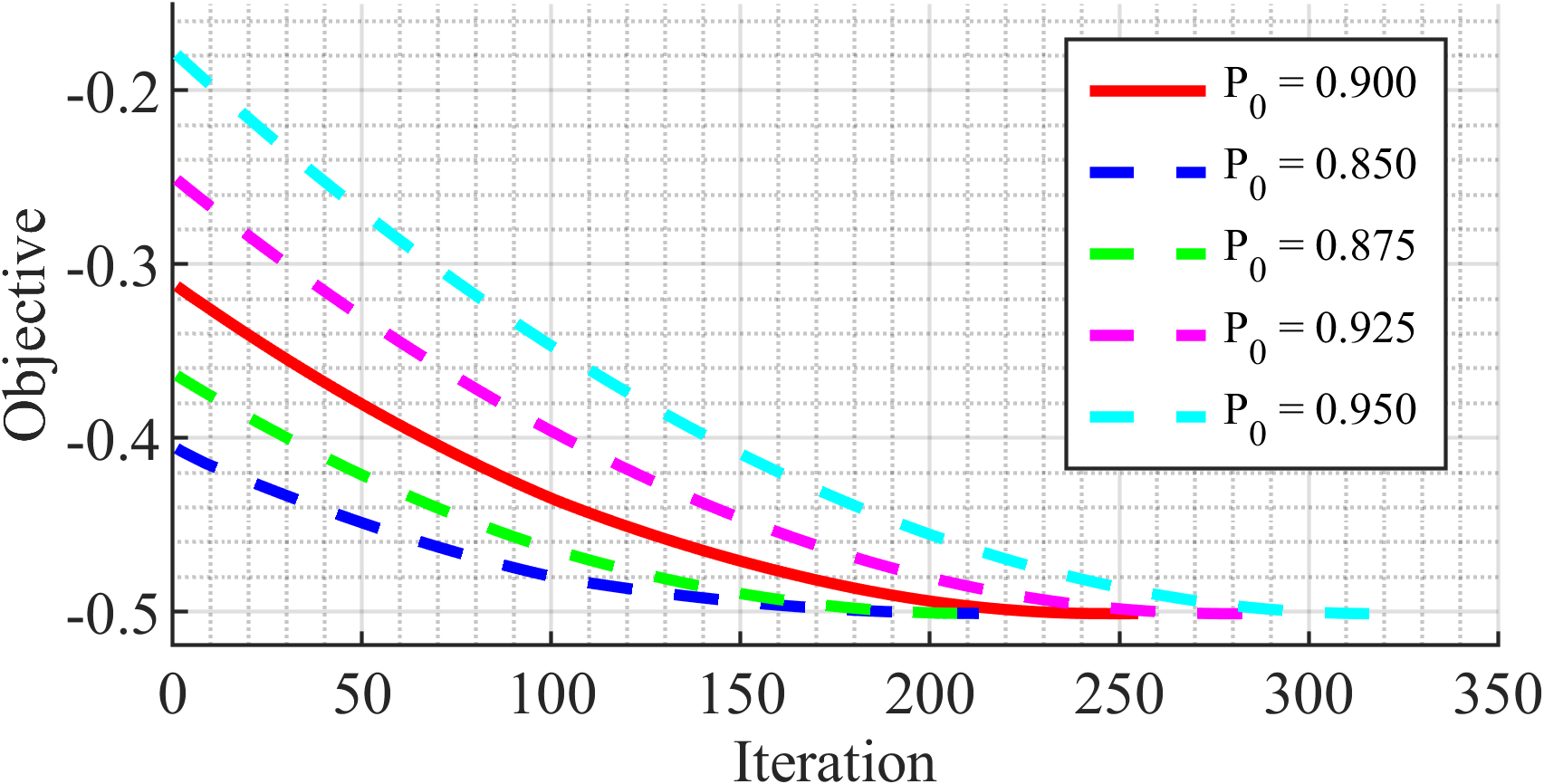}
    \caption{The convergence process of the proposed AO algorithm at $\eta=0.71$, where $P_0$ is the power split ratio in the pilot region.}
    \label{fig:Converge}
\end{figure}

Fig.~\ref{fig:Converge} plots the convergence behavior of the optimization under different power splits applied to the spike-based initialization. The solid curve traces the run that attains the best objective value, and the dashed curves correspond to the other initializations. Fig.~\ref{fig:Converge} shows that proper power initialization markedly accelerates convergence.

\begin{figure*}[!t]
  \centering
  \subfloat[Sensing optimal ($\eta = 0$)%
    \label{fig:power_eta0}]{
      \includegraphics[width=0.31\textwidth]{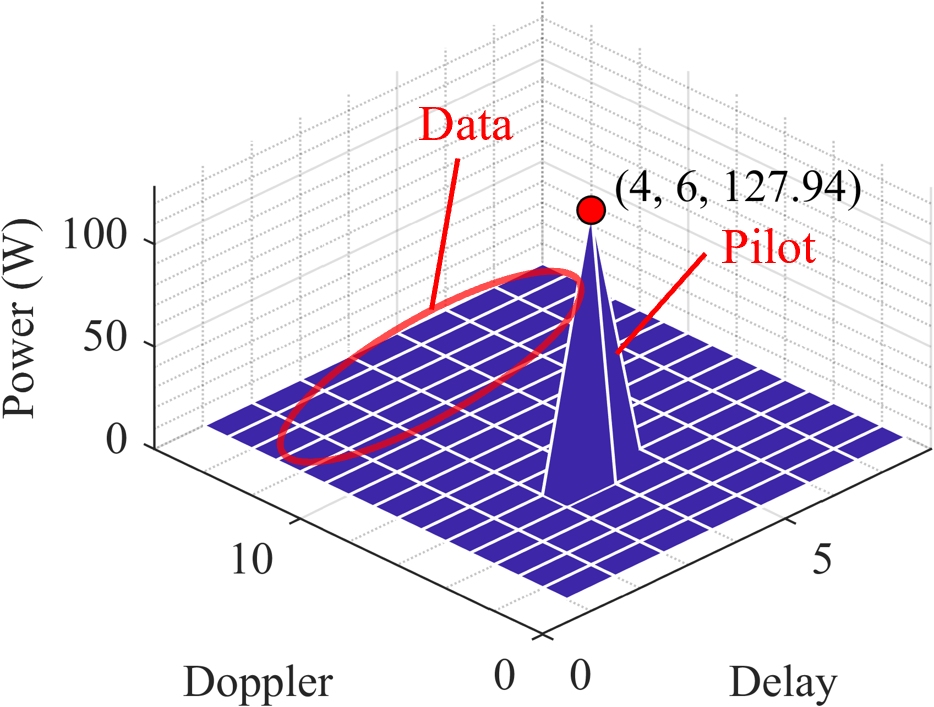}
  }\hfill
  \subfloat[Balanced ($\eta = 0.5$)%
    \label{fig:power_eta05}]{
      \includegraphics[width=0.31\textwidth]{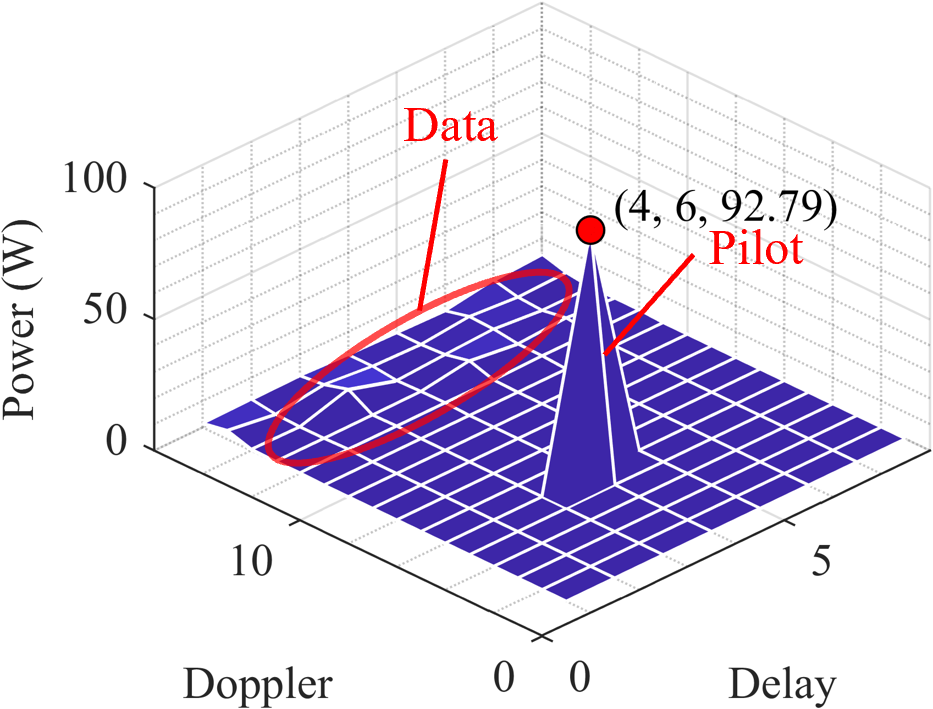}
  }\hfill
  \subfloat[Communication optimal ($\eta = 1$)%
    \label{fig:power_eta1}]{
      \includegraphics[width=0.31\textwidth]{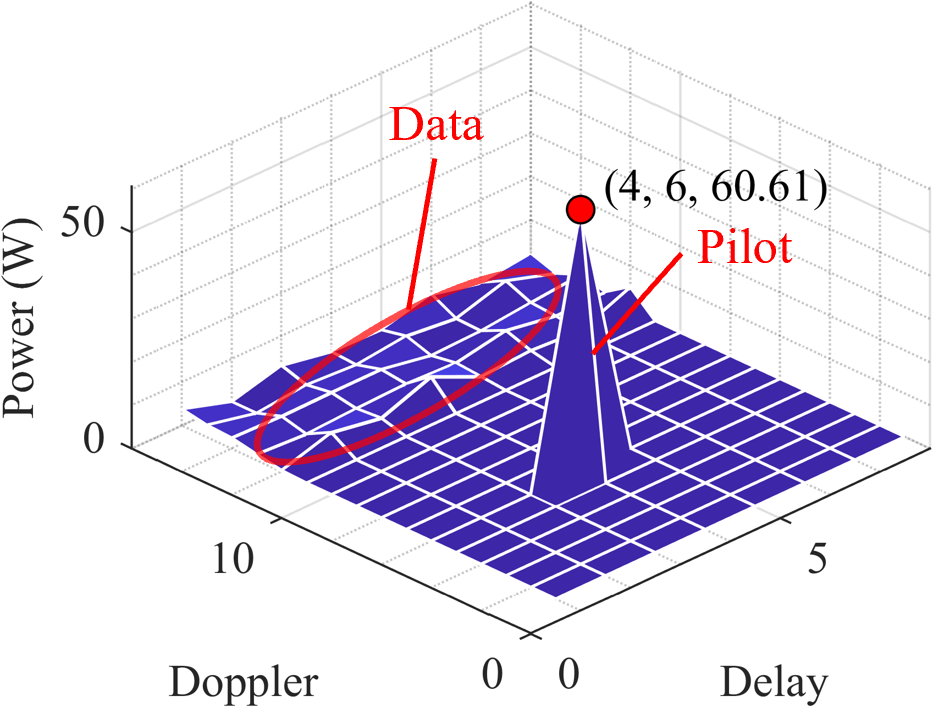}
  }
  \caption{The optimal DD-domain power pattern slices for three different $\eta$ values using the spike arrangement as initialization, with $r_\text{GI}=0.5$ and $r_\text{pilot}=0.375$.}
  \label{fig:power_eta_all}
\end{figure*}

Fig.~\ref{fig:power_eta_all} illustrates the optimal pilot symbols and data-power level in the OTFS DD domain for sensing-centric (i.e., $\eta=0$), intermediate (i.e., $\eta=0.5$), and communication-centric (i.e., $\eta=1$) cases, respectively, given the pilot and data arrangements. Fig.~\ref{fig:power_eta0} demonstrates that a peak-like pilot without power allocated to data symbols in the DD domain is optimal for sensing-centric DFRC. This can be interpreted from two equivalent perspectives: (i) after TF mapping, it occupies the entire TF plane; and (ii) the peak-like DD-domain symbol yields a sharply peaked AF, which is favorable for DD resolution. For communication-centric DFRC, Fig.~\ref{fig:power_eta1} indicates that the peak-like structure is also preserved, minimizing the MSE of channel estimation for OTFS communications. In addition, the optimal results for sensing-centric and communication-centric DFRC not only satisfy the optimality of the two extreme cases in the OTFS DFRC design but also coincide with the optima for pure sensing and pure communication \cite{werf_optimal_2026, chong_DD_2025}, which is a property that may not hold for other waveforms. Fig.~\ref{fig:power_eta05} illustrates an intermediate DFRC case: the optimization shows that OTFS still employs a peak-like pilot, consistent with Fig.~\ref{fig:power_eta0} and Fig.~\ref{fig:power_eta1}; this case differs only in the weighting parameter~$\eta$, which reflects the deterministic–random (DR) trade-off~\cite{zhang_input_2024, liu_deterministic_2023}. Meanwhile, the shared peak-like pilot structure across Fig.~\ref{fig:power_eta0}–Fig.~\ref{fig:power_eta1} reveals an essential consistency between communication channel estimation and sensing, which, in turn, reflects their common underlying physical mechanism.

Fig.~\ref{fig:power_eta_all} also indicates the benefits of OTFS over OFDM in DFRC. Specifically, OTFS enforces multiplexing in the DD domain: data and a peak-like pilot are placed on the DD grid, which after the ISFFT, appear superposed in the TF domain. This structure enables both sensing and communication to exploit the full TF domain without performance degradation, by leveraging the DD domain orthogonality. In contrast, OFDM’s TF multiplexing fixes pilots and data to disjoint regions, preventing either from exploiting the full TF plane for channel estimation or data transmission, thereby limiting TF-optimal performance. Specifically, from a sensing perspective, the DFRC design favors deterministic pilots that would span the TF plane, yet the separation in OFDM prevents pilots from extending into the data region. Conversely, from a communication perspective, DFRC design also demands a large data region that occupies the TF domain, but this is prevented by the pilot region. Furthermore, the resultant OTFS DFRC signal structure mitigates pilot-arrangement complexity—owing to the simplicity of the optimal peak-like pilot’s structure in the DD domain—and thereby facilitates balancing the DFRC trade-off.

% \footnote{Despite these structural advantages of OTFS, a rigorous DFRC performance comparison with OFDM (considering arrangement, pilot symbols, and GI) remains an interesting and open problem.}

\begin{figure*}[!t]
  \centering
  \subfloat[SINR and ISL curves for different $\eta$%
    \label{fig:Opt_SINR_ISL}]{
      \includegraphics[width=0.465\textwidth]{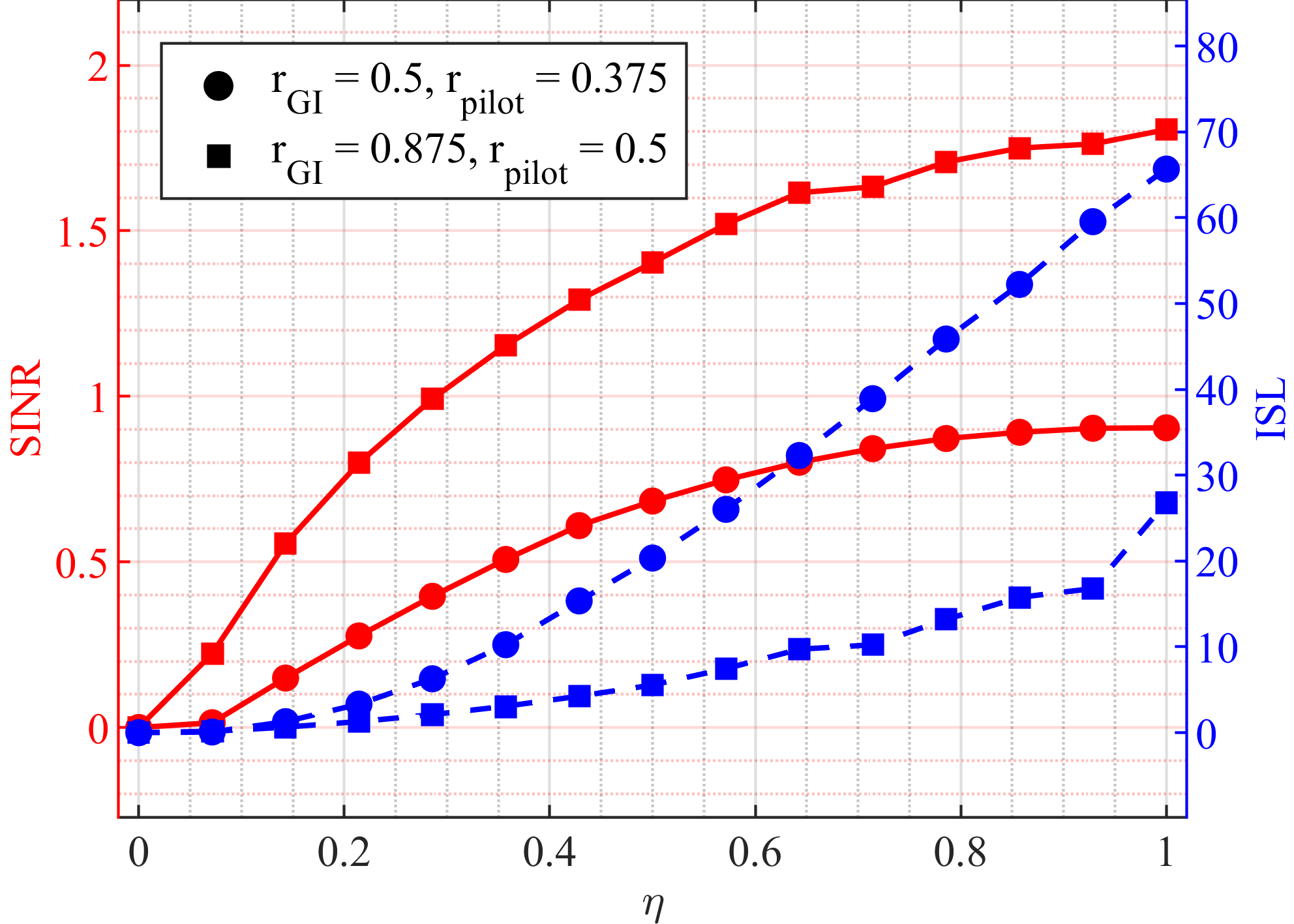}
  }\hfill
  \subfloat[The achievable OTFS DFRC performance region%
    \label{fig:Opt_Region}]{
      \includegraphics[width=0.47\textwidth]{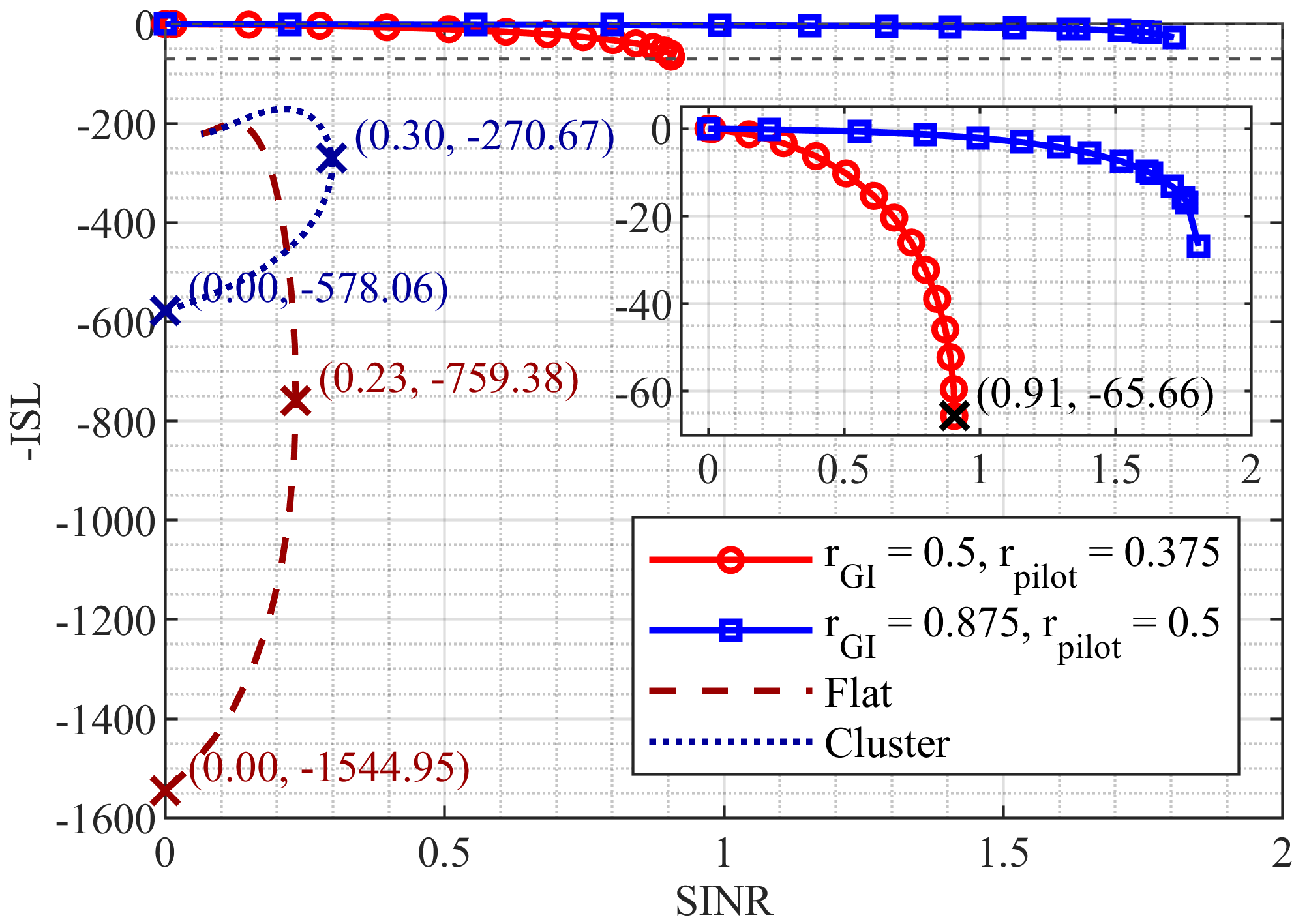}
  }
  \caption{The OTFS DFRC optimization results for different $\eta$ (each metric is normalized to its resultant value at $\eta = 1$ during optimization; for plotting, we show the original scale).}
  \label{fig:Opt_curves}
\end{figure*}
Fig.~\ref{fig:Opt_curves} demonstrates the performance shift as the ISAC weighting factor $\eta$ varies from 0 (sensing-centric) to 1 (communication-centric). In Fig.~\ref{fig:Opt_SINR_ISL}, it could be observed that the SINR increases monotonically from zero to its peak, and the ISL follows the same pattern because both metrics vary with the balance set by $\eta$. The $\text{SINR}$ of the curve with square markers is higher than that of the curve with circle markers because a larger guard interval could concentrate more power into a smaller data-and-pilot region, thereby increasing the SINR. This SINR gain is offset by reduced throughput—a cost that the SINR alone does not show. For the same reason, the ISL of the curve with square markers is lower than that of the curve with circle markers: fewer random data symbols are modulated, so sidelobes are reduced. In Fig.~\ref{fig:Opt_Region}, the DFRC performance region is compared with the flat and cluster arrangements, as illustrated in Fig.~\ref{fig:Scheme_flat} and Fig.~\ref{fig:Scheme_cluster}, shown as solid and dashed lines, respectively. The regions for the flat and cluster arrangements are plotted by varying the power split ratio between data and pilot symbols (dashed lines, progressing from pilot-only at the bottom left to data-only at the top), as shown in Fig.~\ref{fig:Opt_Region}. The optimized design outperforms both other arrangements by enlarging the achievable DFRC performance region, which confirms the effectiveness of the proposed optimization. Compared with the cluster and flat schemes, the proposed signal achieves ISL suppression gains of 9.45 dB and 13.72 dB (when comparing the worst ISL), and SINR gains of 4.82 dB and 5.97 dB (when comparing the best SINR), respectively.

\begin{figure}[h]
  \centering
  \subfloat[Zero-Doppler slice
    \label{fig:AF_Delay}]{
      \includegraphics[width=0.47\linewidth]{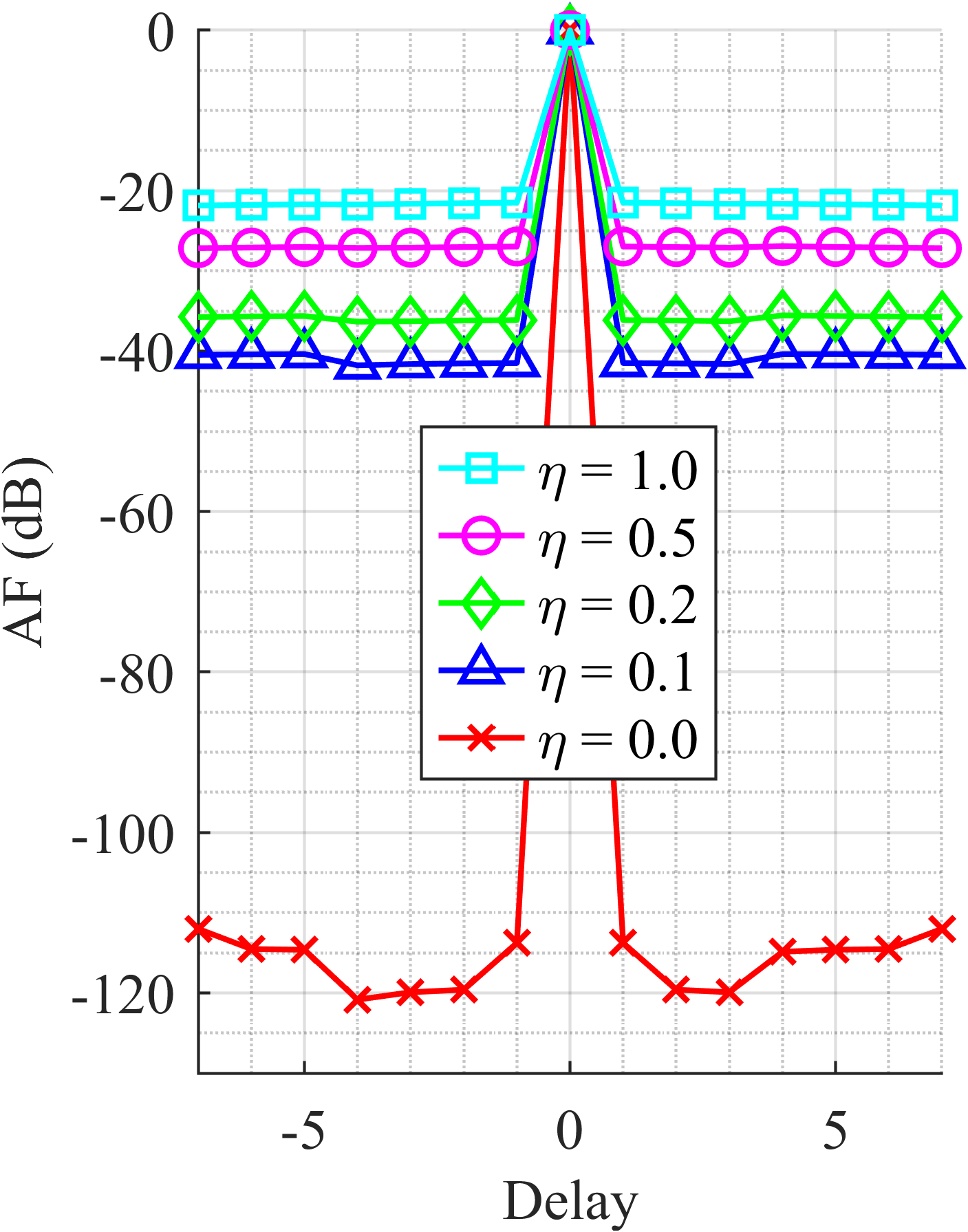}
  }\hfill
  \subfloat[Zero-delay slice
    \label{fig:AF_Doppler}]{
      \includegraphics[width=0.47\linewidth]{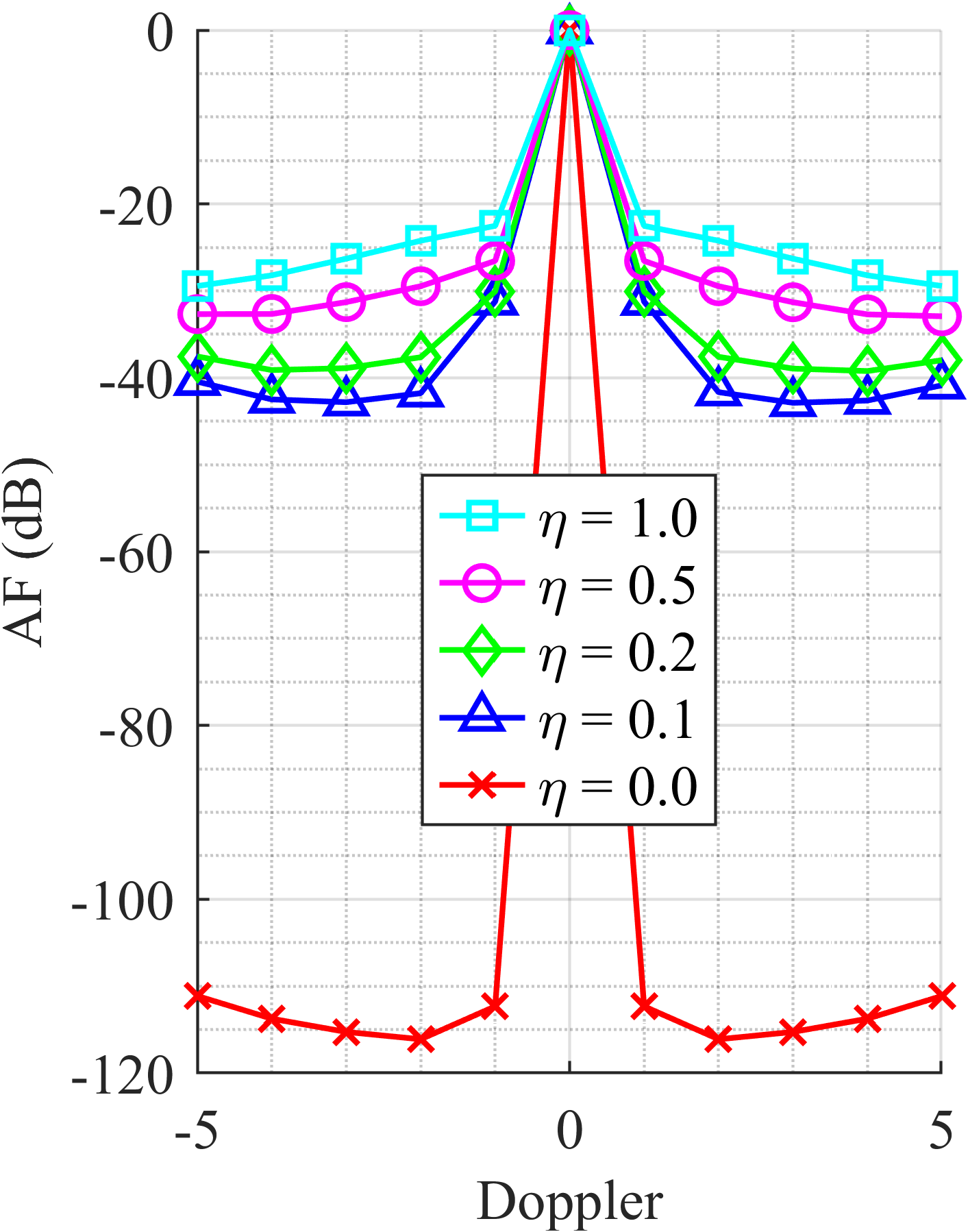}
  }
  \caption{Empirical AFs of optimization results across different ISAC balance parameters, to verify our derived radar metric.}
  \label{fig:AF}
\end{figure}
Fig.~\ref{fig:AF} illustrates the AFs of the optimized OTFS signals for different ISAC balance preferences. It could be seen that the sensing-centric design exhibits a lower sidelobe level than the communication-centric one; as $\eta$ increases, the sidelobes also increase until they reach the communication-centric level. This reflects the DR trade-off: as more power is allocated to the data region, the data symbols—and their randomness—facilitate effective information transmission at the expense of sensing performance.

\begin{figure}[h]
    \centering
    \includegraphics [width=0.80\linewidth]{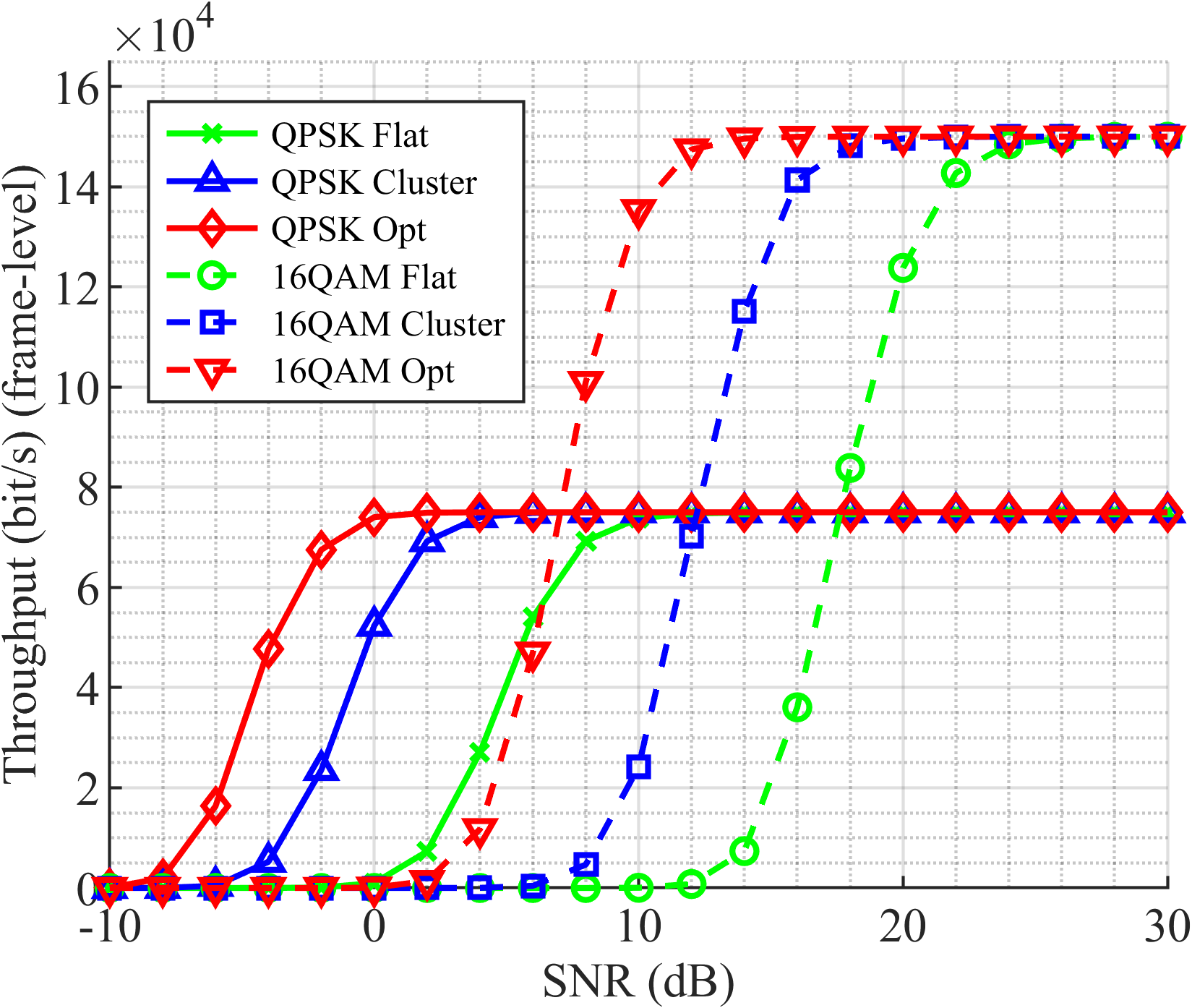}
    \caption{Frame-level throughput versus SNR for OTFS with flat, cluster, and optimized signal arrangements under QPSK and 16QAM. Results are obtained using 1000 channel realizations and 1000 AWGN realizations per channel. The optimized signal adopts $\eta = 1$, $r_\text{GI} = 0.5$, $r_\text{pilot} = 0.375$, with channel parameters $L = 7$ and $Q = 3$.}
    \label{fig:Throughput}
\end{figure}
To further validate the optimal OTFS signal from a communication perspective and to verify the effectiveness of the derived capacity lower bound, we evaluate its frame-level throughput performance, computed as 
\begin{align}
    R(\gamma)=\frac{K_d \log_2 M_{\text{order}}\Delta f}{N}\bigl(1-\mathrm{BER}(\gamma)\bigr)^{K_d \log_2 M_{\text{order}}},
\end{align}
where $M_{\text{order}}$ denotes the modulation order and $\gamma$ is the SNR, as shown in Fig.~\ref{fig:Throughput}. It can be observed that the optimized signal consistently achieves the highest throughput over the entire SNR range, followed by the cluster arrangement and then the flat arrangement. For a given arrangement, the QPSK curves lie to the left of their 16QAM counterparts, reflecting the improved robustness of QPSK to noise, whereas 16QAM attains a higher throughput ceiling owing to its larger modulation order\footnote{Although derived for Gaussian signaling, the capacity lower bound still captures the impact of imperfect CSI for practical constellations. Explicit extensions of this lower bound to finite constellations (e.g., PSK, QAM) are beyond the scope of this paper and are left for future work.}. The fact that the optimized signal reaches the throughput ceiling at the lowest SNR implies the lowest BER and channel estimation error, thereby further confirming the validity of the proposed communication metric.

\section{Conclusion}
\label{sec:conclu}
This paper investigated the design of the DFRC signal for OTFS modulation in ISAC systems, presenting a uniform framework for signaling design across pilots and data payload. We first derived the performance metrics for OTFS DFRC: (1) a channel capacity lower bound, which included channel estimation error and whose dominant term, SINR, was used as the communication metric; (2) the ISL for sensing, which considered the random data symbols and deterministic pilot symbols. We leveraged the above analytical metrics to optimize jointly the pilot signaling and data payload power. Our simulation and analysis showed that OTFS achieved a relatively unified DD-domain design for DFRC that mitigated the common trade-off between sensing and communications: a deterministic, peak-like pilot formed the common structure in the DD domain, facilitating sensing and channel estimation. Relative to conventional OTFS signals, the proposed signal expanded the sensing–communication performance region, achieving at least a 9.45 dB ISL suppression and a 4.82 dB SINR gain.

% if have a single appendix:
%\appendix[Proof of the Zonklar Equations]
% or
%\appendix  % for no appendix heading
% do not use \section anymore after \appendix, only \section*
% is possibly needed

% use appendices with more than one appendix
% then use \section to start each appendix
% you must declare a \section before using any
% \subsection or using \label (\appendices by itself
% starts a section numbered zero.)
%

\appendix
\label{app:channel capacity lower bound}
 Based on the DD domain pilot observation in (\ref{equ:pilot_Omega_h})
 \begin{align}
    \mathbf{y}_\text{DD,p} &= \bm{\Omega}_\text{DD,p}\mathbf{h} + \mathbf{n}_\text{p}, \notag
 \end{align}
the LMMSE weighting matrix could be expressed as
\begin{align}
    \mathbf{W} &= \left( {\bm{\Omega}_{\text{DD,p}}^H} \mathbf{C}_{\mathbf{n}_{\text{p}}}^{-1} \bm{\Omega}_{\text{DD,p}} + \mathbf{C}_{\mathbf{h}}^{-1} \right)^{-1} {\bm{\Omega}_{\text{DD,p}}^H} \mathbf{C}_{\mathbf{n}_{\text{p}}}^{-1} \\
    &= \frac{\sigma_{\mathbf{h}}^2}{\sigma_{\mathbf{n}}^2}  \left( \frac{\sigma_{\mathbf{h}}^2}{\sigma_{\mathbf{n}}^2} \bm{\Omega}_{\text{DD,p}}^H\bm{\Omega}_{\text{DD,p}} + \mathbf{I}_{K_h} \right)^{-1}{\bm{\Omega}_{\text{DD,p}}^H}\\
    &= \frac{\sigma_{\mathbf{h}}^2}{\sigma_{\mathbf{n}}^2} {\bm{\Omega}_{\text{DD,p}}^H} \left( \frac{\sigma_{\mathbf{h}}^2}{\sigma_{\mathbf{n}}^2} \bm{\Omega}_{\text{DD,p}} \bm{\Omega}_{\text{DD,p}}^H + \mathbf{I}_{R_p} \right)^{-1}.
\end{align}
Then, we could rewrite the covariance matrix of the estimated channel vector as follows
\begin{align}
    \mathbf{C}_{\mathbf{\widehat{h}}} &= \varepsilon \Big\{ 
    \mathbf{W} \big( \bm{\Omega}_{\text{DD,p}} \mathbf{h} + \mathbf{n}_{\text{p}} \big) 
    \big( \bm{\Omega}_{\text{DD,p}} \mathbf{h} + \mathbf{n}_{\text{p}} \big)^H 
    \mathbf{W}^H \Big\} \\
    &= 
    \mathbf{W} \big( \sigma_{\mathbf{h}}^2 \bm{\Omega}_{\text{DD,p}} \bm{\Omega}_{\text{DD,p}}^H 
    + \sigma_{\mathbf{n}}^2 \mathbf{I}_{R_p} \big) \mathbf{W}^H  \\
    &=
    \sigma_{\mathbf{h}}^2 \frac{\sigma_{\mathbf{h}}^2}{\sigma_{\mathbf{n}}^2} 
    \bm{\Omega}_{\text{DD,p}}^H 
    \Big( \frac{\sigma_{\mathbf{h}}^2}{\sigma_{\mathbf{n}}^2} 
    \bm{\Omega}_{\text{DD,p}} \bm{\Omega}_{\text{DD,p}}^H 
    + \mathbf{I}_{R_p} \Big)^{-1} \bm{\Omega}_{\text{DD,p}}.
\end{align}
Invoking the Woodbury identity yields the compact form
\begin{align}
    \mathbf{C}_{\mathbf{\widehat{h}}}&= \sigma_{\mathbf{h}}^2 \Big[ 
    \mathbf{I}_{K_h} - \Big( 
    \mathbf{I}_{K_h} + \frac{\sigma_{\mathbf{h}}^2}{\sigma_{\mathbf{n}}^2} 
    \bm{\Omega}_{\text{DD,p}}^H \bm{\Omega}_{\text{DD,p}} 
    \Big)^{-1} 
    \Big].
\end{align}
In addition, $\mathbf{C}_{\mathbf{h}} = \sigma_{\mathbf{h}}^2 \mathbf{I}_{K_h}$. Therefore,
\begin{align}
    {\mathbf{C}_{{\epsilon}}} = {\mathbf{C}_{{\mathbf{h}}}} - {\mathbf{C}_{{\mathbf{\widehat h}}}} = \sigma_{\mathbf{h}}^2 \left(\mathbf{I}_{K_{h}} + \frac{\sigma_{\mathbf{h}}^2}{\sigma_{\mathbf{n}}^2}{{\bm{\Omega}}^H _\text{DD,p}}{\bm{\Omega}} _\text{DD,p}\right)^{-1} \notag
\end{align}
which matches (\ref{equ:C_eps}) in the main text.

% Can use something like this to put references on a page
% by themselves when using endfloat and the caption.
% \ifCLASSOPTIONcaptionsoff
%   \newpage
% \fi

\bibliographystyle{IEEEtran}
\bibliography{bibtex/bib/IEEEabrv,bibtex/bib/IEEEexample}

\end{document}